\documentclass[aps,prd,amsfonts,amssymb, amsmath, showpacs, showkeys, a4paper, nofootinbib, superscriptaddress, 11pt, raggedbottom]{revtex4-2}
\usepackage{mathtools, latexsym}\usepackage{graphicx, float, subfig}
\usepackage[normalem]{ulem}
\usepackage[colorlinks=true, allcolors=blue]{hyperref}
\usepackage{natbib}
\usepackage{makecell}
\usepackage{soul}

\newcommand{\RE}{\mathrm{Re}}

\begin{document}
\title{Using the Pericentre Precession of LAGEOS II to Constrain Quadratically Coupled Ultralight Dark Matter}
\author{Clare Burrage}
\email{clare.burrage@nottingham.ac.uk}
\author{Angus Macdonald}
\email{angus.macdonald@nottingham.ac.uk}
\affiliation{School of Physics and Astronomy, University of Nottingham,
University Park, Nottingham NG7 2RD, UK}
\author{Elisa Todarello}
\email{elisa.todarello@berkeley.edu}
\affiliation{Leinweber Institute for Theoretical Physics, University of California, Berkeley, CA 94720, U.S.A.}
\affiliation{Theoretical Physics Group, Lawrence Berkeley National Laboratory, Berkeley, CA 94720, U.S.A.}
\affiliation{Universit\`a degli Studi di Torino, via P. Giuria 1, I--10125 Torino, Italy}

\begin{abstract}
    It has been proposed that feebly-interacting ultralight scalars may constitute the dark matter content of the universe. {Models describing the interactions of a dark matter scalar with Standard Model fields may feature quadratic interactions at leading order, such that the scalar acquires an effective mass in the neighbourhood of a classical matter distribution}. The effect of their introduction is to provide effective shifts in fundamental constants of physics, as well as to mediate scalar `fifth forces' between test bodies. We here demonstrate how these fifth forces can result in pericentre precession in the orbital motion of satellites around the Earth. We apply this to the measured pericentre precession of the LAGEOS II experiment, constraining the mass, and couplings to the light Standard Model fields, of a quadratically coupled ultra-light dark matter scalar. We observe such an experiment to be effective in constraining parameter space at strong couplings, where existing constraints from satellite and tabletop level experiments break down.
\end{abstract}
\maketitle

\newpage
\tableofcontents

\newpage
\section{Introduction}
It is currently understood that an undetected Dark Matter constitutes approximately $25\%$ of the energy balance of the universe \cite{Planck:2018vyg}. Ultra-light scalar fields are a promising candidate for dark matter \cite{Hu:2000ke, Hui:2021tkt, Antypas:2022asj}. These constitute a class of particle with masses $\ll 1\rm{~eV}$, and necessarily feeble interactions with ordinary, observable matter. Similar particles are ubiquitous in models attempting to extend and resolve problems within the Standard Model, including the QCD axion, as a solution to the Strong CP Problem \cite{Wilczek:1977pj, Weinberg:1977ma, Sikivie:1983ip}, dilatons and moduli from the compactification of extra dimensions in string theory \cite{Damour:1994ya,Damour:1994zq,Kaplan:2000hh,Cicoli:2023opf}, and attempts to modify General Relativity (GR)  \cite{McDonald:1991xm, McDonald:1993ma,Burrage:2018dvt,Clifton:2011jh,Joyce:2014kja}.

In accordance with the properties of the galactic dark matter halo \cite{Drukier:1986tm} and the small ascribed field masses, these scalars are expected to have simultaneously high number densities and long de Broglie wavelengths in the solar neighbourhood, leading to high state occupation numbers \cite{Ferreira:2020fam}. As such, it is possible to locally describe ultra-light scalar dark matter as a classical condensate. Whilst non-gravitational interactions between dark matter and ordinary matter have yet to be observed, in the absence of a principle forbidding interactions with matter, we must assume they exist. It is shown in Refs.~\cite{Damour:2010rp, Damour:2010rm} that if such a scalar is introduced into the Standard Model through couplings to light fields, then scalar-induced effective shifts in the fundamental constants of physics should be observed as a consequence. In turn, these lead to phenomena such as scalar-mediated fifth forces, apparent equivalence principle violation, and oscillations in atomic energy transitions. These effects have been the basis of many attempts to search for, and constrain the properties of, ultralight scalar dark matter using tabletop and satellite scale experiments. These include, but are not limited to: tests of the universality of free-fall \cite{Touboul:2017grn,MICROSCOPE:2019jix}, precision measurements of transitions in atomic clocks \cite{Banks:2024sli, 
Brzeminski:2026rox,Filzinger:2023qqh,Derevianko:2013oaa}, Cavendish-like torsion balance experiments \cite{Schlamminger:2007ht, Wagner:2012ui, Luo:2024ocg}, and atom interferometry \cite{MAGIS-100:2021etm,Bertoldi:2021rqk,AEDGE:2019nxb,Abend:2023jxv,Badurina:2019hst,Antypas:2019qji}.

The effectiveness of these searches will depend on the nature of couplings to matter featured in the scalar theory. Whilst a model featuring linear interactions to leading order might be most simple, there has been recent interest in models where quadratic terms dominate scalar couplings to Standard Model fields at low energies \cite{Brzeminski:2026rox, Bouley:2022eer, Hees:2018fpg, Bouley:2022eer,Sibiryakov:2020eir, Banerjee:2022sqg}. Indeed, the physics of such models is relevant to axions \cite{Bauer:2023czj,Grossman:2025cov,Banerjee:2025dlo}, scalars which may couple to the Standard Model through a Symmetric Higgs Portal \cite{OConnell:2006rsp,Patt:2006fw,Burgess:2000yq,Arcadi:2019lka}, and other scalar models where a mirror symmetry forbids linear interaction terms \cite{Delaunay:2025pho}. 

The effect of introducing quadratic couplings into a scalar theory is that the effective mass of the field shifts in the vicinity of matter. For positive coupling constants, this leads to the suppression of the field in the vicinity of massive objects. For negative coupling constants, this can lead to both suppression and enhancement of the scalar amplitude around matter, as well as possible tachyonic instability in the scalar theory. {In previous work \cite{Burrage:2025grx,Day:2023mkb}, it has been noted} that the suppressive effects induced by a matter environment surrounding an experiment, such as a vacuum chamber, laboratory room, or any other such experimental housing, could significantly weaken existing constraints at strong couplings. This was particularly observed for positive coupling constants between the scalar and matter. However, in \cite{Burrage:2025grx} we noted that experiments tracking the motion of satellites should still be able to place constraints on strong couplings, since satellites are largely isolated from other objects.

In this paper, we demonstrate the constraining power of satellite tracking experiments, using the ``Laser Geodynamic Satellite-2'' (LAGEOS II) experiment; a spherical satellite, informally compared to a `disco ball' due to its retro-reflectors, and which sits in a particularly stable orbit around the Earth, which has enabled precise tracking of its motion through laser ranging techniques. Data from this experiment has been previously applied in tests of GR and modified gravity \cite{LUCCHESI:2003234, Lucchesi:2014uza, Lucchesi:2010zzb, Iorio:2001eg, Feleppa:2025vop, Ciufolini:2019cux}. 

In Sec.~\ref{sec:quad_coupling_dm}, we introduce a basic model for quadratically coupled dark matter. In Sec.~\ref{sec:field_solutions}, we then give an outline describing field solutions around massive bodies; in particular, solid spheres and binary systems of spheres. The latter of these we use to approximate the field profile around the Earth-LAGEOS II system. In Sec.~\ref{sec:precession_calculation}, we then determine the pericentre precession rate of a satellite's orbit expected from fifth forces in a quadratically coupled scalar theory. In Sec.~\ref{sec:constraints}, we present new constraints on quadratically coupled ultra-light dark matter (Fig.~\ref{fig:Precession Constraints}), based on measurement of the anomalous peri-centre precession of LAGEOS II, which we compare against existing constraints. Finally, in Sec.~\ref{sec:approximations}, we move to a discussion of approximations made in our calculations. Throughout this work, unless otherwise specified, we will work in natural units: $\hbar=c=1$, and assume the `mostly minus' metric convention where relevant. 

\section{Quadratically Coupled Dark Matter}
\label{sec:quad_coupling_dm}
This work introduces dark matter through the addition of a massive scalar field $\phi$ to the Standard Model. We presume that, via a mirror symmetry or otherwise, interactions which are linear in $\phi$ are forbidden. Following the notation of Refs. \cite{Damour:2010rm, Damour:2010rp, Hees:2018fpg}, scalar interactions with light Standard Model fields are described through an effective theory:
\begin{equation}
\begin{split}
    \mathcal{L} \supseteq  \frac{1}{2}&(\partial\phi)^2-\frac{1}{2}m^2\phi^2\\&+\frac{(\kappa\phi)^2}{2}\left[\frac{d_e^{(2)}}{4}F_{\mu\nu}F^{\mu\nu}-\frac{d_g^{(2)}\beta_3}{2g_3}F^A_{\mu\nu}F^{A~{\mu\nu}}-\sum_{f={u,d,e}}(d_{m_f}^{(2)}+\gamma_{m_f}d_g^{(2)})m_f\bar{\psi}_i{\psi}_i\right] \\&+\mathcal{L}_{SM}\;,
\end{split}
    \label{eq:scalar_theory}
\end{equation}
where $\kappa=\sqrt{4\pi G}$, $F$ and $F^{A}$ are respectively the photon and gluon field strength tensors, $\beta_3$ is the QCD beta function, $\gamma_{m_f}$ are the quark anomalous dimensions, $\psi_i$ are Dirac spinors for the light fermions, and $\mathcal{L}_{SM}$ describes Standard Model physics.  $d_e^{(2)},~d_g^{(2)}~d_{m_f}^{(2)}$ are the so-called ``dilaton coefficients", and are unknown dimensionless constants in the scalar theory. It can be demonstrated that such couplings yield effective shifts in the fundamental constants of physics - particularly the QCD mass scale $\Lambda_3$, the fine-structure constant $\alpha_{EM}$, and the masses of the light fermionic fields $m_f$:
\begin{equation}
    \begin{split}
        \Lambda_3(\phi)&=\Lambda_3\left(1+d_g^{(2)}\frac{(\kappa\phi)^2}{2}\right)\;,\\
        \alpha_{EM}(\phi)&=\alpha_{EM}\left(1+d_e^{(2)}\frac{(\kappa\phi)^2}{2}\right)\;, \\
        m_f(\phi)&=m_f\left(1+d_{m_f}^{(2)}\frac{(\kappa\phi)^2}{2}\right)\;.\\
    \end{split}
    \label{eq:constant_shifts}
\end{equation}
This leads to shifts in the masses of composite objects. For atoms of mass $m_A$:
\begin{equation}
    m_A(\phi)=m_A\left(1+\frac{\alpha_A\phi^2}{2}\right)
    \label{eq:mass_shift}
\end{equation}
where the coupling parameter $\alpha_A$ is parametrised as a linear superposition of the dilaton coefficients:
\begin{equation}
    \alpha_A =  \kappa^2\Big[d_g^{(2)}+Q^A_{\hat{m}}(d_{\hat{m}}^{(2)}-d_g^{(2)})+Q^A_{\delta m}(d_{\delta m}^{(2)}-d_g^{(2)})+Q^A_{m_e}(d_{m_e}^{(2)}-d_g^{(2)})+Q^A_{e}d_{e} \Big]\;.
    \label{eq:Alpha_Definition}
\end{equation}
The dilaton charges $Q^A_i$ are calculated semi-empirically in terms of atomic number, and mass number  \cite{Damour:2010rm,Damour:2010rp}, and $\alpha_A$ is composition dependent. The composite coefficients $d_{\delta m}^{(2)},~d_{\hat{m}}^{(2)}$ are determined from the coefficients $d_{m_u},~d_{m_u}$:
\begin{equation}
    \begin{split}
    d_{\hat{m}}^{(2)} &= \frac{ m_{u} d_{m_u} + m_{d} d_{m_d}}{m_{u} + m_{d}}\;,\\
    d_{\delta m}^{(2)} &= \frac{ m_{u} d_{m_u} - m_{d} d_{m_d}}{m_{u} - m_{d}}\;.
\end{split}
\end{equation}
From such principles, one can determine the action describing the scalar in the vicinity of some classical, non-relativistic matter density $\rho_A(x)$:
\begin{equation}
    S = \int d^4x \left(\frac{1}{2}(\partial\phi)^2-\frac{1}{2}m^2\phi^2-\frac{1}{2}\rho_A(x)\alpha_A\phi^2\right)\;,
    \label{eq:scalar_theory_alpha}
\end{equation}
whereby the scalar equation of motion is given:
\begin{equation}
    \Box\phi+(m^2+\rho_A\alpha_A)\phi=0\;.
    \label{eq:scalar_eom}
\end{equation}
It is clear that the scalar develops an effective squared-mass $m_{eff}^2=m^2 + \rho_A\alpha_A$ when in the vicinity of matter. Whilst the values of $d_i^{(2)}$, and hence $\alpha_A$, may in principle be either positive or negative, modelling scalar behaviour around matter is challenging in the latter case. This arises from the possibility for the field to become tachyonic inside matter for sufficiently negative couplings, or inside regions of particularly dense ordinary matter, such that: $\alpha_A\rho_A\ll -m^2$. As such, in this work, we will focus our attention on $\alpha_A>0$. Additionally, we may restrict our analysis to parameter space where the values of atomic masses and the fundamental constants do not experience significant temporal or spatial modulation due to the dark matter scalar, which are inconsistent with day-to-day observations. Combined, these correspond to:
\begin{equation}
    \begin{split}
        d_g^{(2)}\frac{(\kappa\phi)^2}{2}\ll1\;,\\
        d_e^{(2)}\frac{(\kappa\phi)^2}{2}\ll1\;,\\
        d_{m_f}^{(2)}\frac{(\kappa\phi)^2}{2}\ll1\;,\\
        \alpha_A\frac{\phi^2}{2}\ll1\;.
    \end{split}
    \label{eq:theory_breakdown}
\end{equation}
These conditions also ensure that the effective field theory in Eq.~\eqref{eq:scalar_theory} is valid, and that interaction terms of quartic or higher order $\sim \mathcal{O}\left[(d_i^{(2)}(\kappa\phi)^2)^2\right]$, can be safely neglected.
\section{Analytical Field Solutions Around Matter Distributions}
\label{sec:field_solutions}
\subsection{Homogeneous Vacuum Solution}
Given the long lengthscales associated with the galactic halo, we assume that the distribution of dark matter should be broadly isotropic and homogeneous, except in the immediate vicinity of compact sources of ordinary matter. Describing this space assuming the baryonic matter density $\rho=0$, then solutions of Eq.~\eqref{eq:scalar_eom} take the form:
\begin{equation}
    \phi=\phi_\infty\cos(mt+\delta)\;,
    \label{eq:homogeneous}
\end{equation}
where we will hence absorb the unknown phase factor $\delta$ into the time co-ordinate $t$. The implications of relaxing this description of the field on our results are discussed in Sec.~\ref{sec:DM_wind}. The amplitude of oscillations is set by the local dark matter density $\rho_{DM}\simeq0.4 \rm{~GeV ~cm}^{-3}$ \cite{deSalas:2020hbh}.
\begin{equation}
    \phi_\infty=\sqrt{\frac{2\rho_{DM}}{m^2}}
\end{equation}

\subsection{Field Solutions Around Spherical Bodies}
\label{sec:spherical_solutions}
In this work, we are particularly interested in the scalar field profiles around both the Earth and the LAGEOS II satellite, and the consequent scalar fifth forces. Conveniently, both bodies are very well described as spheres. Earth may be modelled as an oblate spheroid with eccentricity $~0.0818$ \cite{WGS84} (see further discussion in Sec.~\ref{sec:approximations}) and LAGEOS II is described as a sphere of $\sim60 \rm{cm}$ diameter, spherical up to sub-millimetre precision, for the benefit of laser ranging. \cite{LAGEOS:comp_tech_report, LAGEOS:Techreport}%\footnote{we here refer to section 5.2.2 of a technical report relating to the LAGEOS I satellite - the predecessor to LAGEOS II. However, since from \cite{LAGEOS:Techreport} we observe that an objective in the design of LAGEOS II was to develop a satellite "as nearly identical to LAGEOS I as possible" - then it is expected that LAGEOS II should be spherical up to similar precision.} . Hence, we will discuss field solutions around spheres, and binary systems thereof.
\subsubsection{Around a Uniform Sphere}
\label{sec:uniform_sphere}
The simplest body around which we may determine field solutions is a sphere of radius $R_S$ and constant density $\rho_S$, giving a density profile:
\begin{equation}
    \rho(x)=\rho_S\Theta(R_S-r) \;.
\end{equation}
The scalar profile around this source mass is then determined by solving Eq.~\eqref{eq:scalar_eom} with this matter density - requiring field continuity and differentiability, and using Eq.~\eqref{eq:homogeneous} to describe the scalar profile at $r=\infty$. Solutions have the form:
\begin{equation}
    \phi_S(r,t)=\phi_\infty\cos(mt)\begin{dcases}
        \frac{1}{\cosh(k_SR_S)}\frac{\sinh(k_Sr)}{k_Sr} & r<R_S\\
        1+\frac{q(k_SR_S)R_S}{r} & r>R_S \;,
    \end{dcases}
    \label{eq:spherical_solution}
\end{equation}
where we define:
\begin{equation}
    q(x)=\frac{\tanh(x)-x}{x}\;.
\end{equation}
From Eq.~\eqref{eq:spherical_solution}, we observe field solutions inside the sphere to have a characteristic decay scale: $k_S=\sqrt{\rho_S\alpha_S}$, defining a skin depth $\sim1/k_S$ for the penetration of the field into the sphere. We note that this field profile has two key regimes of behaviour as the coupling constant is varied: a strong coupling limit $k_SR_S\gg1$, where the field amplitude inside the sphere is exponentially suppressed as $k_S$ increases, and a weak coupling regime, $k_SR_S\ll1$, where the field asymptotes to the background solution $\phi(r,t)\rightarrow\phi_\infty\cos(mt)$ both inside and outside the source, as $k_S\rightarrow0$. 
\subsubsection{Around a Layered Sphere}
\label{sec:Layered Sphere}
Whilst the simplicity of the preceding solutions may make describing both the Earth and LAGEOS II as uniform spheres appealing, we can more accurately approximate each as having a layered structure. LAGEOS II has a brass core\footnote{The core of the satellite has a cylindrical form. We comment on this in Sec.~\ref{sec:modelling_LAGEOS}.} surrounded by an aluminium shell, whilst Earth's structure may be broadly divided into an iron-dominated core and silicate mantle. Both bodies can therefore be described via the density distribution:
\begin{equation}
        \rho_l(r)=\begin{dcases}
            \rho_l^{i}  & r< R_l^{i}\\
            \rho_l^{e}  & R_l^{i}<r< R_l^{e}\\
            0           & r>R_l^{e}
        \end{dcases}
\end{equation}
Where $\rho_l^{i},~\rho_l^{e}$ are the densities of the interior and exterior regions of the sphere's layered structure, bounded by the radii $R_l^{i},~R_l^{e}$. Solutions around a body with similar structure have previously been derived in Ref.~\cite{Hees:2018fpg}. Defining corresponding coupling constants $\alpha$ such that $k_l^{i}=\sqrt{\alpha_l^{i}\rho_l^{i}}$, $k_l^{e}=\sqrt{\alpha_l^{e}\rho_l^{e}}$, and again assuming the sphere to sit on a homogeneous background, field solutions may be derived with the form:
\begin{equation}
    \phi_{l}(r,t)=\phi_\infty\cos(mt)\begin{dcases}
        A_{l}^{int}\frac{\sinh(k^{i}_lr)}{k_l^{i}r} & r<R_l^{i}\\
        A_{l}^{ext}\frac{\sinh(k^{e}_l r)}{k^{e}_l r} + B_{l}^{ext}\frac{\cosh(k^{e}_lr)}{k^{e}_lr} & R^{i}_l<r<R^{e}_l\\
        1+\frac{B_{l}^{out}}{r} &r>R^{e}_l
    \end{dcases}
    \label{eq:layered_sphere_solutions}
\end{equation}
where the coefficients $A_{l},~B_{l}$ are again determined from continuity and differentiability conditions, and are given in Appendix~\ref{app:field_coeffs}. As with the uniform sphere case, we observe from these coefficients that the field inside matter is strongly screened when the skin depth of the field inside both the exterior and interior regions becomes comparable to the length scale of each region: $k^i_lR^i_l\sim1$, $k^e_l(R^e_l-R^i_l)\sim1$. In the limit of strong coupling between the scalar field and matter $k^{e/i}_l\rightarrow\infty$, we note that the field profile tends to a `maximally screened' state - identical to the uniform sphere case for $R_l^e=R_S$:
\begin{equation}
    \lim_{k^{e/i}_l\rightarrow\infty}\phi(r)=\phi_\infty\cos(mt)\begin{dcases}
    0 & r<R^e_l\\
    1-\frac{R_l^e}{r} & r>R^e_l
    \end{dcases}
\end{equation}

Throughout this work, we will use the solutions in Eq.~(\ref{eq:layered_sphere_solutions}) to define the field profile around the Earth. The densities of matter in each region have been chosen in accordance with the `Preliminary Reference Earth Model' \cite{DZIEWONSKI1981297}, and are outlined in Table~\ref{tab:spheres_properties} alongside properties of the LAGEOS II satellite.
\subsection{Solutions in Binary Systems}
\label{sec:solutions_near_earth}
In order to derive scalar-mediated, Earth-LAGEOS II fifth forces in Sec.~\ref{sec:precession_calculation}, we will require a description of the scalar profile around the satellite, incorporating the influence of both bodies. Solving the scalar equations of motion both analytically and exactly is expected to be challenging. As such, we here present an approximate method for finding analytical solutions around LAGEOS II, in the vicinity of Earth. In particular, we describe the scalar profile around the satellite as perturbing a background scalar profile provided by the Earth. We assume that LAGEOS II significantly perturbs the field profile around the Earth in a sufficiently small region of space, such that a linearised description of the Earth's field may be used.

We consider the Earth and LAGEOS II as spheres with radii $R_\oplus,~R_L$, and with layered density structures. We will use two co-ordinate bases: $\vec{r}_\oplus=\vec{r}_\oplus(r_\oplus,\theta_\oplus,\psi_\oplus)$, about the centre of the Earth, and $\vec{r}_L=\vec{r}_L(r_L,\theta_L,\psi_L)$ centred on LAGEOS II. We additionally define $\vec{r}$ as the position of the centre of LAGEOS II, relative to the centre of the Earth, such that $\vec{r}_\oplus=\vec{r}+\vec{r}_L$. We choose $\theta_L=\pi$ to always lie on the line-of-centres between the two bodies, and $\theta_\oplus=\frac{\pi}{2}$ to lie on the orbital plane of LAGEOS II. We additionally define $h=r-R_\oplus$ as the height of the centre of the satellite above the surface of the Earth. 

The field profile around the Earth is assumed to take the form $\phi_\oplus=\phi_l(r_\oplus,t;k_\oplus^{i},k_\oplus^{e})$, as described in Eq.~\eqref{eq:layered_sphere_solutions}. We linearise this around the position of LAGEOS:
\begin{equation}
    \phi_\oplus(\vec{r}_\oplus=\vec{r}+\vec{r}_L)\approx\phi_\infty\cos(mt)\big(a_0(r)+a_1(r) r_L\cos(\theta_L)\big)\;,
    \label{eq:Earth_Taylor_Expansion}
\end{equation}
where the coefficients $a_0, a_1$ are determined:
\begin{equation}
\begin{split}
    a_0(r)&=1+\frac{B^{out}_\oplus}{r}\\
    a_1(r)&=-\frac{B^{out}_\oplus}{r^2}\;.
\end{split}
\label{eq:a0a1_definition}
\end{equation}
Conservatively, this linearisation is limited to describing the field in a neighbourhood $r_L\ll h$, with the changing effective mass of the field at the Earth's surface being responsible its breakdown outside this region. 

We now consider how LAGEOS II perturbs this linearised field. We note that in Eq.~\eqref{eq:layered_sphere_solutions}, the field around a layered sphere radius $R$ decays to the homogeneous background field profile as $R_l^e/r\times f(k_l^i,k_l^e)$ where $f$ is some factor: $0\leq f\leq 1$. Given that LAGEOS II is positioned at a height $h\gg R_L$, then the satellite's contribution to the field profile should have decayed significantly before reaching the boundary $r_L\sim h$. Consequently, we determine the field profile around LAGEOS II by solving Eq.~\eqref{eq:scalar_eom}, whilst using Eq.~\eqref{eq:Earth_Taylor_Expansion} as an approximate boundary condition in the limit $r_L\rightarrow\infty$. These solutions take the form:
\begin{equation}
    \begin{split}
    &\phi(r_L)=\phi_\infty\cos(mt) \times\\
        &~~~~~~\begin{dcases}
             A_{L,0}^{int}\frac{\sinh(k_L^{i}r_L)}{k_L^{i} r_L} 
             + A_{L,1}^{int}\left(\frac{\cosh(k_L^{i}r_L)}{k_L^{i} r_L}-\frac{\sinh(k_L^{i}r_L)}{(k_L^{i} r_L)^2}\right)\cos(\theta_L) & r_L<R_L^{i}\\
            \begin{split}
                &A_{L,0}^{ext}\frac{\sinh(k_L^{e}r_L)}{k_L^{e} r_L} + A_{L,1}^{ext}\left(\frac{\cosh(k_L^{e}r_L)}{k_L^{e} r_L}-\frac{\sinh(k_L^{e}r_L)}{(k_L^{e} r_L)^2}\right)\cos(\theta_L) \\
                &+B_{L,0}^{ext}\frac{\cosh(k_L^{e}r_L)}{k_L^{e} r_L} + B_{L,1}^{ext}\left(\frac{\sinh(k_L^{e}r_L)}{k_L^{e} r_L}-\frac{\cosh(k_L^{e}r_L)}{(k_L^{e} r_L)^2}\right)\cos(\theta_L)
            \end{split}& R_L^{int}<r_L<R_L^{ext}\\
            \left(a_0 + \frac{B_{L,0}^{out}}{r_L}\right) +\left(a_1 r_L+\frac{B_{L,1}^{out}}{r_L^2} \right)\cos( \theta_L) &  R_L^{ext}<r_L\ll h\;.
        \end{dcases}
    \end{split}
    \label{eq:linearised_solution}
\end{equation}
The coefficients $A_L, B_L$ are given in Appendix~\ref{app:field_coeffs}. We further discuss the validity of these solutions in Sec.~\ref{sec:liniarised_approx_breakdown}.

\section{Scalar-Induced Orbital Precession in Newtonian Gravity}
\label{sec:precession_calculation}
We now calculate the pericentre precession arising from scalar-mediated fifth forces on one body orbiting another. We shall later use this, in conjunction with precision observations of LAGEOS II in Ref.~\cite{Lucchesi:2014uza}, to constrain parameter space for a quadratically coupled scalar. The pericentre precession observed in Ref.~\cite{Lucchesi:2014uza} is consistent with that predicted by GR, but the measurement precision still allows for some small anomalous precession, which constrains fifth forces. Since both GR and fifth force corrections to the motion of LAGEOS II are necessarily small, we expect GR corrections to fifth forces themselves to be negligible. As such, in deriving the scalar-induced pericentre precession, we work within a Newtonian framework. Given the hierarchy of masses $M_\oplus/M_L\sim\mathcal{O}[10^{22}]$, we safely assume that the barycentre of the orbital system lies at the centre of the Earth.

The scalar-mediated fifth force experienced by LAGEOS II, is determined to quadratic order in $\phi$ by:
\begin{equation}
    \vec{F}_5=-\int d^3x \;\alpha_L \rho_L(x)\frac{\nabla\phi^2}{2}\;.
\end{equation}
Inserting the solutions for $\phi$ in Eq.~\eqref{eq:linearised_solution}, we observe this force to be written in the form:
\begin{equation}
\begin{split}
    \vec{F}_5=\eta(k_{L}^{i},k^e_{L})\vec{F}_5^{limit} 
\end{split}
\end{equation}
Where $\vec{F}_5^{limit}=-4\pi a_0 a_1 R_{L}^e \phi_\infty^2\cos^2(mt)\hat{r}$ is the force in the strong coupling limit $k_L^{i/e}\rightarrow\infty$, and is independent of the satellite's internal density and composition. The coefficient $\eta$ is defined as:
\begin{equation}
    \eta(k_L^i,k_L^e)=\left(1-\frac{k_L^eR_L^i\tanh(k_L^iR_L^i)+k_L^iR_L^i\tanh(k_L^e(R_L^e-R_L^i))}{k_L^eR_L^e(k_L^iR_L^i+k_L^eR_L^i\tanh(k_L^iR_L^i)\tanh(k_L^e(R_L^e-R_L^i))}\right)\;,
\end{equation}
and satisfies $0\leq\eta\leq1$ for $k_L^i,k_L^e\in\mathbb{R}$. The total force acting on LAGEOS II is then:
\begin{equation}
    M_L[\phi]\ddot{\vec{r}}=-\frac{GM_L[\phi]M_\oplus[\phi]}{r^2}\hat{r}+ \vec{F}_5
    \label{eq:total_force_1}
\end{equation}
where the separation of centres $\vec{r}$ has been promoted to a dynamical variable, though we still choose to fix $\theta=\frac{\pi}{2}$ to the orbital plane of LAGEOS II. The masses $M_i[\phi]=M_i+\delta M_i\cos^2(mt)$\footnote{For a macroscopic body $i$, such as LAGEOS II or Earth, the mass shift $\delta M_i$ can be calculated: $\delta M_i\cos^2(mt)=\int_{V_i}d^3x\rho(x)\tfrac{\alpha(x)\phi^2(x,t)}{2}$.} incorporate scalar-induced corrections as in Eq.~\eqref{eq:mass_shift}. From the expansion of $\ddot{\vec{r}}$:
\begin{equation}
    \ddot{\vec{r}}=\left(\ddot{r}-r\dot{\psi}^2\right)\hat{r}+\frac{1}{r}\frac{d}{dt}(r^2\dot{\psi})\hat{\psi}\;,
\end{equation}
from which, the absence of $\hat{\psi}$ terms in Eq.~\eqref{eq:total_force_1} yields the typical angular conservation:
\begin{equation}
    \dot{\psi}=\frac{l}{r^2}\;.
    \label{eq:angular_velocity}
\end{equation}
Through the substitution of $a_0, ~a_1$ from Eq.~\eqref{eq:a0a1_definition} into Eq.~\eqref{eq:total_force_1}, the equation of for the radial motion of the satellite may be written \:
\begin{equation}
\begin{split}
    \ddot{r}=&\frac{-GM_\oplus}{r^2}\left(1+\frac{\delta M_\oplus}{M_\oplus}\cos^2(mt)-\frac{4\pi R_L^e\phi_\infty^2 B_\oplus^{out}}{GM_\oplus M_L}\eta\cos^2(mt)\right) \\&+ \frac{l^2}{r^3}\left(1+\frac{4\pi R_L^e\phi_\infty^2 {B_\oplus^{out}}^2}{l^2M_L}\eta\cos^2(mt)\right)
\end{split}
    \label{eq:radial_acceleration_1}
\end{equation}
where we neglect $F_5\delta M_L \sim \mathcal{O}[(\alpha\phi^2)^2]$ contributions to the motion of LAGEOS II. We will assume that LAGEOS II has an orbital period $\tau_L$ satisfying $m\tau_L\gg1$, such that we may average over oscillatory terms. Importantly, given the satellite's $\sim 13,300~\rm{s}$ orbital period \cite{Lucchesi:2014uza}, this assumption restricts our analysis to the regime $m\gtrsim10^{-20}~\rm{eV}$. We additionally define a new set of constants:
\begin{align}
    \tilde{G}&=G\left(1+\frac{\delta M_\oplus}{2M_\oplus}-\frac{2\pi R_L^e\phi_\infty^2 B_\oplus^{out}}{G M_\oplus M_L}\eta\right)
    \label{eq:G_tilde}
    \\
    \tilde{l}&=l\left(1+\frac{2\pi R_L^e\phi_\infty^2 {B_\oplus^{out}}^2}{l^2 M_L}\eta\right)^\frac{1}{2}
    \label{eq:l_tilde}
\end{align}
such that Eq.~\eqref{eq:radial_acceleration_1} takes the form:
\begin{equation}
    \ddot{r}=-\frac{\tilde{G} M_\oplus}{r^2}+\frac{\tilde{l}^2}{r^3}
    \label{eq:radial_accelleration_2}
\end{equation}
% Using the substitution: $u=\frac{1}{r}$, and using Eq.~\eqref{eq:angular_velocity}, Eq.~\eqref{eq:radial_accelleration_2} may be put into the form:
% \begin{equation}
%     \frac{d^2u}{d\psi^2}=-\frac{\tilde{l}^2}{l^2}\left(u-\frac{GM}{\tilde{l}^2}\right)
% \end{equation}
% where it is used that:
% \begin{equation}
% \begin{split}
% \frac{dr}{du}&=-\frac{1}{u^2}=-r^2\\
% \frac{dr}{dt}&=\frac{d\psi}{dt}\frac{dr}{du}\frac{d u}{d\psi}=(lu^2)\left(-\frac{1}{u^2}\right)\frac{d u}{d\psi}
% \end{split}
% \end{equation}
% We may then solve for $u(\psi)$, and hence $r(\psi)$:
Using Eq.~\eqref{eq:angular_velocity}, this may be solved exactly for $r$ as a function of the angular co-ordinate $\psi$. Mathematically, this is achieved by: first, multiplying both sides of Eq.~\eqref{eq:radial_accelleration_2} by a factor $\dot{r}$ and integrating with respect to time:
\begin{equation}
    \frac{1}{2}\dot{r}^2=\frac{\tilde{G}M_\oplus}{r}-\frac{\tilde{l}^2}{2r^2}
    \label{eq:calc_step_1}
\end{equation}
second, dividing both sides of Eq.~\eqref{eq:calc_step_1} by $\dot{\psi}^2$:
\begin{equation}
    \frac{1}{2}\left(\frac{d r}{d\psi}\right)^2=\left(\frac{\tilde{G}M_\oplus}{r}-\frac{\tilde{l}^2}{2r^2}\right)\frac{r^4}{l^2}
\end{equation}
third, rewriting the equation of motion in terms of the reciprocal of the radial co-ordinate $u=\frac{1}{r}$
\begin{equation}
    \frac{1}{2}\left(\frac{d u}{d\psi}\right)^2\frac{1}{u^4}=\left(\tilde{G}M_\oplus u-\frac{1}{2}\tilde{l}^2u^2\right)\frac{1}{l^2u^4}
\end{equation}
and finally solving the resulting equation of motion in $u$:
\begin{equation}
    \frac{d^2u}{d\psi^2}=\frac{\tilde{G}M_\oplus}{l^2}-\frac{\tilde{l}^2}{l^2}u=-\frac{\tilde{l}^2}{l^2}\left(u-\frac{\tilde{G}M_\oplus}{\tilde{l}^2}\right)\;.
\end{equation}
Consequently, under the effects of Newtonian gravity, and a fifth force mediated by a quadratically coupled scalar, LAGEOS II follows an orbital trajectory:
\begin{equation}
    r(\psi)=\frac{\tilde{l}^2}{\tilde{G}M_\oplus}\frac{1}{1+e\cos\left(\frac{\tilde{l}}{l}\psi+\delta\right)}\;.
\end{equation}
Here, $e>0,~\delta$ are constants of integration, describing the eccentricity and initial phase of the orbit respectively. For convenience, we absorb $\delta$ into the angular co-ordinate $\psi$. %The satellite then passes through the pericentre of it's orbit where $({\tilde{l}}/{l})\psi=2\pi n$, for integer $n$ \AM{Opinions if this sentence needed?}. 
Defining the shift in the pericentre of the orbit: $\Delta\omega(t)=\left(\psi(t) -\tfrac{\tilde{l}}{l}\psi(t)\right)$, scalar fifth forces predict a pericentre precession rate over many cycles:
\begin{equation}
    \Delta\dot{\omega}_\phi=\frac{2\pi}{\tau_{L}}\left(1-\frac{\tilde{l}}{{l}}\right)\approx-\frac{2\pi^2 R_{L}^e\phi_\infty^2 {B_\oplus^{out}}^2}{l^2M_L\tau_{L}}\eta\;.
    \label{eq:pericentre_precession}
\end{equation}
Notation has been chosen in accordance with the orbital elements of celestial mechanics, in which $\omega$ is the `argument of pericentre', and $\Delta\omega(t)$ is its shift over a duration $t$. We have here implicitly assumed: $\left|1-\tfrac{\tilde{l}}{{l}}\right|\ll1$. LAGEOS II's orbit has an orbital period $\tau_L~\simeq13,300~\rm{s}$, semi-major axis $=12.162\times10^3~\rm{km}\simeq 1.9~R_\oplus$, and eccentricity $0.014$ \cite{Lucchesi:2010zzb}.
\begin{table}[]
    \centering
    \begin{tabular}{c|c|c|c|c}
                    &  $\rho^i$ ($\rm{kg~m}^{-3}$) & $\rho^e$ ($\rm{kg~m}^{-3})$& $R^i$ ($\rm{m}$) & $R^e$ ($\rm{m}$)\\ \hline
        Earth       &  $11000$ & $4400$ & $3.5\times10^6$ & $6.4\times10^6$\\ 
        LAGEOS II   & $8280$ & $2740$ & $0.177$ & $0.3$  
    \end{tabular}
    \caption{Properties of the Earth and LAGEOS II satellite, approximated as 'layered spheres', used in our analysis.}
    \label{tab:spheres_properties}
\end{table}

\section{Results}\label{sec:constraints}
\begin{figure}\centering
    \subfloat[]{\includegraphics[width=.5\linewidth]{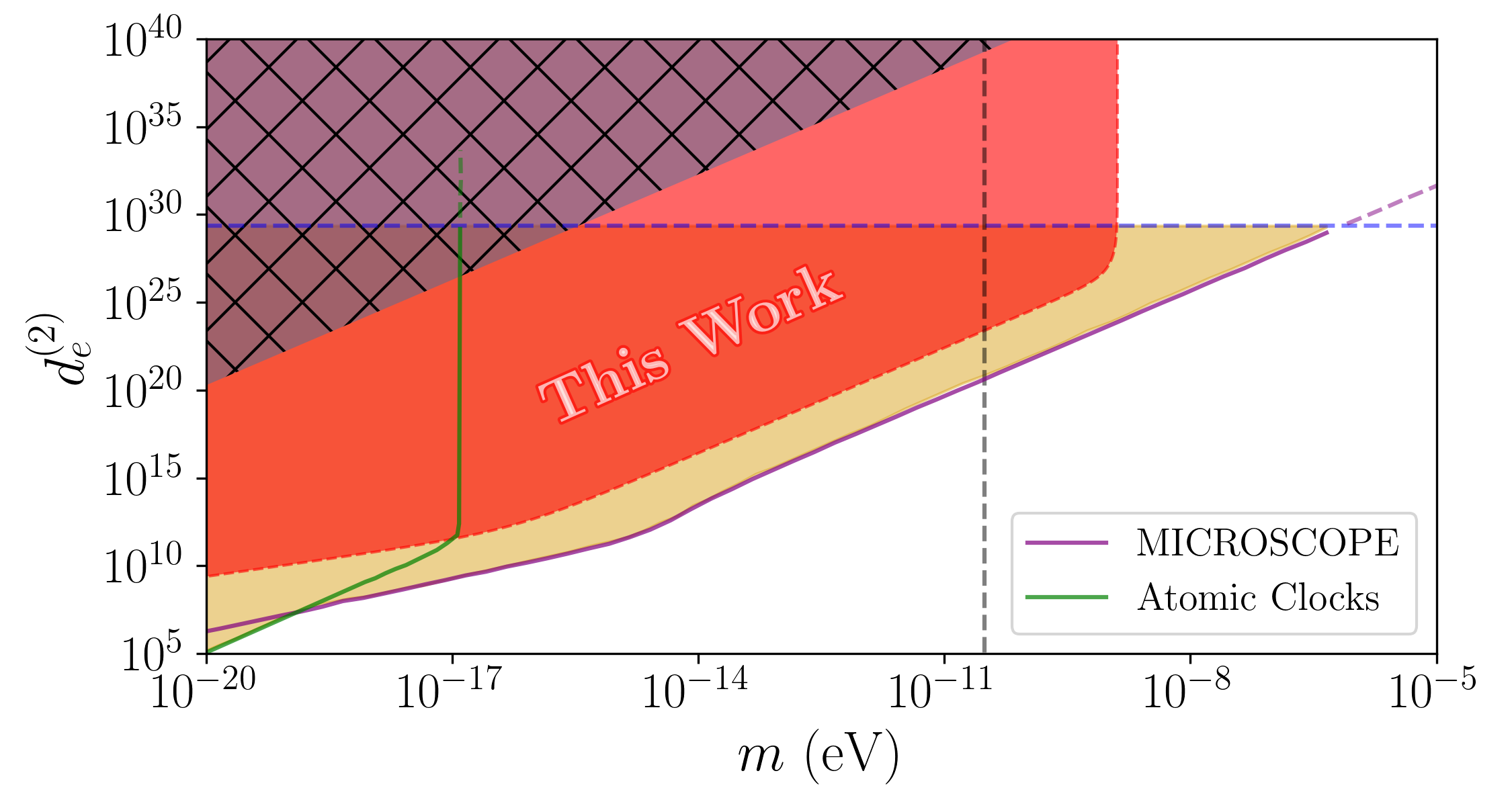}\label{subfig:a}}\hfill
    \subfloat[]{\includegraphics[width=.5\linewidth]{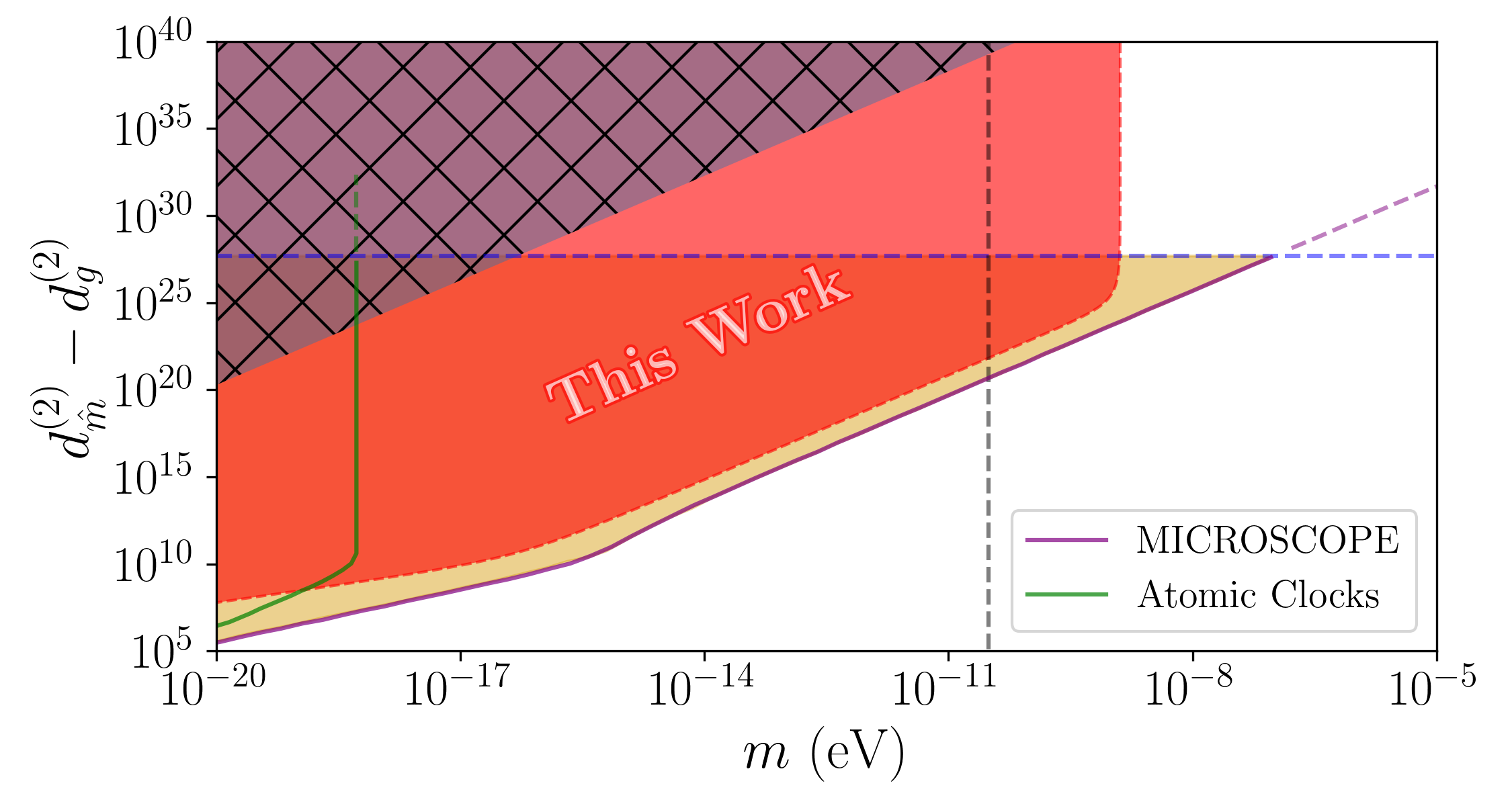}\label{subfig:b}}\par 
    \subfloat[]{\includegraphics[width=.5\linewidth]{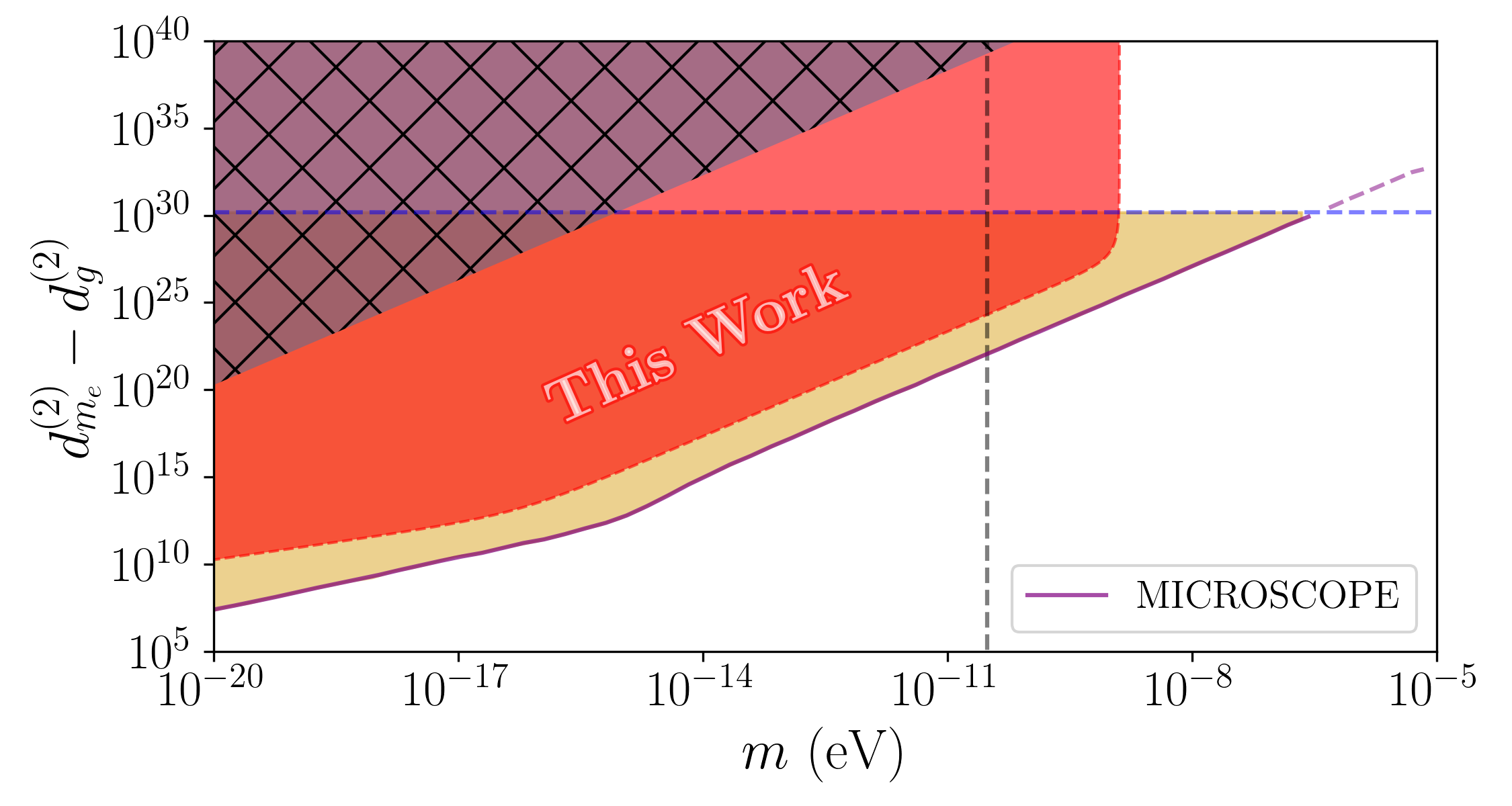}\label{subfig:c}}
    \subfloat[]{\includegraphics[width=.5\linewidth]{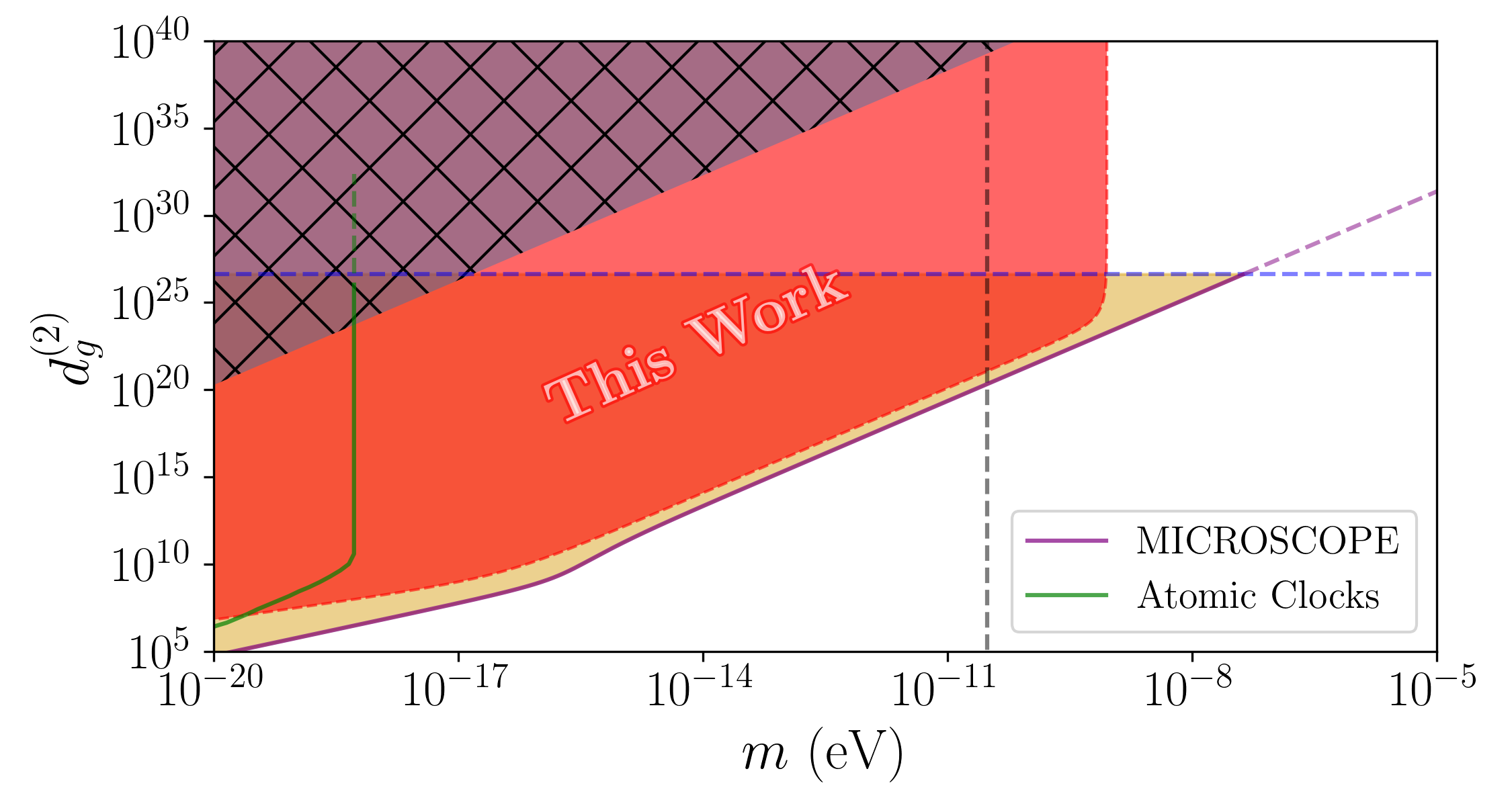}\label{subfig:d}}
    \caption{Constraints on the mass: $m$ of the scalar field, and the maximum reach of combinations of the dilaton coefficients: $d_e^{(2)},~d_{m_e}^{(2)}-d_{g}^{(2)},~d_{\hat{m}}^{(2)}-d_{g}^{(2)}$ from the precession of the perihelion of the LAGEOS II satellite (\textit{red}), where parameter space predicting $\Delta\dot{\omega} _{\phi}\lesssim-181 \rm{~mas~yr^{-1}}$ is excluded. In \textit{yellow}, we display previous constraints from Ref.~\cite{Hees:2018fpg}, showing bounds from the MICROSCOPE experiment \cite{Touboul:2017grn, MICROSCOPE:2019jix} (\textit{purple}), and from atomic clocks \cite{Hees:2016gop, VanTilburg:2015oza} (\textit{green}). The \textit{blue dashed line} gives an approximate threshold at which we expect constraints from Ref.~\cite{Hees:2018fpg}, particularly MICROSCOPE, to be insufficiently conservative \cite{Burrage:2025grx}. The vertical \textit{black dashed line} indicates the mass scale above which dark matter wind effects could complicate constraints - those from both previous authors and work herein. The \textit{dark hatched} region indicates where we expect at least one of the conditions in Eq.~\eqref{eq:theory_breakdown} to be violated, such that the effective theory in Eq.~\eqref{eq:scalar_theory} either is invalidated and/or predicts unphysically large modulation of fundamental constants.}
    \label{fig:Precession Constraints}
\end{figure}
With the result in Eq.~\eqref{eq:pericentre_precession}, we may use experimental measurements of the pericentre precession of LAGEOS II to constrain the parameter space of quadratically coupled ultralight scalars. We do so, utilising Ref.~\cite{Lucchesi:2014uza}. This work presents bounds on the anomalous precession of the satellite in the format:
\begin{equation}
    \Delta\dot{\omega}_{an}\approx3294.95 \times(-1.2\times10^{-4}\pm2.10\times10^{-3}\pm2.54\times10^{-2}) ~\rm{mas~yr^{-1}} \;.
    \label{eq:anomalous_precession}
\end{equation}
In this datum, $3294.95~\rm{mas~yr}^{-1}$ is the expected relativistic contribution to the precession of the satellite's pericentre. The former of the two errors derives from a combined second-order polynomial/Fourier fit to LAGEOS II's residual pericentre precession over time - spanning a $~13$ year period. $\Delta\dot{\omega}$ is determined from the term in the fit which is linear in time. The latter error derives from systematic errors in the models used by the authors to determine the expected relativistic precession of LAGEOS II. For further details, we refer the reader to Ref.~\cite{Lucchesi:2014uza}. Constraints derived from this result are determined by rejecting any combinations of parameters, i.e. the scalar mass $m_i$ and dilaton coefficients $d^{(2)}_e,~d^{(2)}_{\hat{m}},~d^{(2)}_{\delta m}, d^{(2)}_{m_e},d^{(2)}_g$, for which the theoretical pericentre of LAGEOS II in Eq.~\eqref{eq:pericentre_precession} disagrees with the anomalous value at the $95\%$ confidence level, i.e. we exclude parameter space for which
\begin{equation}
    |\Delta\dot{\omega}_\phi-\Delta\dot{\omega}_{an}|>2\sigma\;.
\end{equation}
The total $2\sigma$ error is determined using both errors in Eq.~\eqref{eq:anomalous_precession}, though the result is dominated by the systematic contribution. This ultimately leads to an exclusion of parameter space for which $\Delta\dot{\omega}_\phi\lesssim-180~\rm{mas~yr^{-1}}$.

Our constraints are presented in Fig.~\ref{fig:Precession Constraints}. We constrain a range of field masses consistent with the condition $m\gg1/\tau_L \sim5\times10^{-20}\rm{~eV}$, on which our determination of the perihelion precession of LAGEOS II in Sec.~\ref{sec:precession_calculation} relies. Whilst the mass range shown in Fig.~\ref{fig:Precession Constraints} does contain field masses for which this condition is only barely satisfied, it is at larger masses where the experiment places constraints in the otherwise unconstrained region above the blue dashed threshold in each plot that is of most interest to us. We additionally indicate a possible upper bound on the mass range across which constraints are valid, shown by the black dashed line at $m\simeq\tfrac{10^{3}}{R_\oplus}\simeq 3\times10^{-11} {\rm{eV}}$ in each panel which indicates where a possible dark matter wind, deriving from the relative motion of the Earth through the galactic Dark Matter Halo, could result in a break-down of the homogeneous boundary conditions assumed in Eq.~\eqref{eq:homogeneous}. We note that this mass limit should additionally apply to previous constraints such as those shown for comparison in Fig.~\ref{fig:Precession Constraints}. A brief analysis of the impact of these effects is presented in Sec.~\ref{sec:DM_wind}.

Constraints are determined through a principle of `maximum reach': the value of all but a single term in Eq.~\eqref{eq:Alpha_Definition} is set to zero, leaving the remaining term to be bounded at the order of magnitude level. The format of Fig.~\ref{fig:Precession Constraints}, particularly the combinations of the dilaton coefficients $d_i$ we constrain in Subfigs.~\ref{subfig:a}-\ref{subfig:c}, is chosen to mirror previous bounds by Hees et al.~in Ref.~\cite{Hees:2018fpg}, which we show for comparison. These originate from searches for scalar-induced frequency shifts in atomic clocks \cite{VanTilburg:2015oza, Hees:2016gop} and bounds on fifth forces \cite{Schlamminger:2007ht, Wagner:2012ui, Touboul:2017grn, MICROSCOPE:2019jix}. We choose to distinguish these, particularly highlighting constraints from the `MICROSCOPE' experiment \cite{Touboul:2017grn, MICROSCOPE:2019jix}, which are the strongest over the range of field masses shown. In Subfig.~\ref{subfig:d}, we have also chosen to show constraints on $d^{(2)}_g$ in isolation, not presented in Ref.~\cite{Hees:2018fpg}. Whilst we observe that constraints deriving from the LAGEOS II pericentre precession are generally competitive with previous constraints, we particularly observe this for constraints on the isolated $d_g^{(2)}$. This is owed to the fact that measurements from MICROSCOPE are of differential forces between test masses, which scale as the difference in coupling $\Delta\alpha$ between them. Hence, they are insensitive to universal contributions to the coupling $\alpha$, such as the comparatively large isolated $d_g^{(2)}$ term in Eq.~\eqref{eq:Alpha_Definition}. In this case, developments in the modelling of Earth's gravity, which dominate the error in Eq.~\eqref{eq:anomalous_precession}, but also longer-term, more precise tracking of the motion of LAGEOS II and future satellite experiments, could see them place the tightest bounds within the shown mass-range.

The shape of the region excluded by pericentre precession can be understood as follows. At weak couplings, as the chosen combinations of $d_i$ increase in magnitude, the excluded region expands to cover greater field masses. This is a consequence of the fifth forces, and subsequently the pericentre precession rate, increasing with $d_i$ and scaling as $1/m^2$. As the couplings $d_i$ increase further, the panels in Fig.~\ref{fig:Precession Constraints} demonstrate two clear changes in the direction of the boundary of the excluded region. The first of these, occurs around the $d_i$ for which $\mathcal{O}[k_\oplus^{i/e}R_\oplus^{i/e}]\sim1$. In this region, the skin depth of the scalar field inside the Earth becomes comparable to the radius of the planet, resulting in a screening of the field in its vicinity. This exposes LAGEOS II to stronger gradients in the scalar field about the Earth, but also to reduced field amplitudes. The net effect is that the fifth forces, and subsequently the rate of pericentre precession, increase less dramatically with the values of the $d_i$, for a fixed $m$, above this threshold. A similar behaviour is observed in the MICROSCOPE constraints. This is expected as both sets of constraints derive from scalar-mediated fifth forces between the Earth and test masses. Continuing to larger couplings, the second and most substantial change in the slope of the boundary corresponds to where $\mathcal{O}[k^{e/i}_LR_L^{e/i}]\sim1$. Above this threshold, LAGEOS II is now expected to significantly screen the field in its vicinity. As noted in Sec.~\eqref{sec:precession_calculation}, this screening behaviour leads to the fifth force between Earth and LAGEOS II reaching a maximum limit. Hence, pericentre precession no longer increases with coupling strength, and the boundary of our excluded region in Fig.~\ref{fig:Precession Constraints} sits at a constant $m$ as the various combinations of the $d_i$ increase in magnitude. Constraints from MICROSCOPE data, from Ref.~\cite{Hees:2018fpg}, do not account for this effect, owing to an approximation of test masses as point particles.

Of particular interest is the region of parameter space above the blue dashed threshold. In this region, previous constraints should be relaxed, due to suppression of the amplitude of the scalar field from matter enclosing each experiment \cite{Burrage:2025grx}. This line approximates the satellite-housing of the MICROSCOPE experiment as an aluminium shell of $O(100~\rm{cm})$ diameter and $O(1{\rm ~cm})$. A similar threshold will exist for other constraints from laboratory-based experiments. Because LAGEOS II sits in an environment which is isolated from other sources of matter, the constraints derived in this work can access this region of parameter space.

\section{Approximations}\label{sec:approximations}
In the following subsections, we give further detail and justification for assumptions made in the preceding calculations. 

\subsection{Modelling LAGEOS II}\label{sec:modelling_LAGEOS}
\begin{table}[]
    \centering
    \begin{tabular}{c|c c c c|}
         Material & $Q_{e}$ & $Q_{\hat{m}}$ & $Q_{m_e}$ & $Q_{\delta m}$  \\ 
         Species & $(\times10^{-3})$ & $(\times10^{-2})$ & $(\times10^{-4})$ & $(\times10^{-4})$\\ \hline\hline
         $^{63.5}_{29}\rm{Cu}$&$2.70$&$8.34$&$2.51$&$1.48$\\
         $^{65.4}_{30}\rm{Zn}$&$2.77$&$8.35$&$2.52$&$1.41$\\
         $^{27}_{13}\rm{Al}$&$1.73$&$8.07$&$2.65$&$0.618$\\
         $^{55.8}_{26}\rm{Fe}$&$2.59$&$8.31$&$2.56$&$1.17$\\
         \hline
         \thead{LAGEOS II Core\\(CuZn39Pb2)} & $2.76$ & $8.35$ & $2.51$ & $1.49$
         \\ \thead{LAGEOS II Shell\\(AlMgSiCu 6170)} & & ~~~~~~~See & Al~~~~~~~ &
         \\\hline \thead{Earth Core\\(Iron)} & & ~~~~~~~See & Fe~~~~~~~ &
         \\ \thead{Earth Mantle\\(Silicate)} & $1.61$ & $7.96$ & $2.74$ & $0.0239$
         
    \end{tabular}
    \caption{Dilaton charges for the dominant atomic species, and subsequently the composite material species, needed to model the Earth, and the LAGEOS II experiment.}
    \label{tab:dilaton_values}
\end{table}
Our modelling of the LAGEOS II satellite is guided by Refs.~\cite{LAGEOS:Techreport, VISCO20161928}. As previously described, the satellite is broadly constructed of a Brass core ($\sim59\%$ Cu, $\sim39\%$ Zn, $\sim2\%$ Pb by mass) and an Aluminium shell ($\sim 96\%$ Al, $\sim4\%$ other by mass). We may reasonably describe this system as a layered sphere, the properties of which are outlined in Table~\ref{tab:spheres_properties}. Deviation from this model is expected to arise from the following features of the satellite: that the core is cylindrical rather than spherical, that running through the diameter of the satellite exists a Copper-Beryllium bolt, with a similar density to its core, and that the satellite has germanium retroreflectors inset into its outer shell. The bolt and retroreflectors constitute only $\sim5\%$ of the total mass of the satellite, and hence their presence is expected to have a minimal impact on the motion of the satellite. Subtle variations in internal structure and composition of test masses can have interesting consequences for constraints based on differential measurements - for example, the difference in scalar fifth force between test bodies of identical mass, but differing internal composition \cite{Burrage:2025grx}. However, since the theoretical pericentre precession of LAGEOS II is derived from the \textit{total} scalar fifth force acting upon it, we expect that the bulk properties of the satellite should be most important. Hence, fully accounting for the cylindricity of the core should constitute only minor corrections to its pericentre precession rate.    

Coupling constants $\alpha_A$ for the relevant atomic species in LAGEOS II and the Earth are determined from the dilaton coefficients using Eq.~\eqref{eq:Alpha_Definition}. The resultant $\alpha_C$ for composite materials, such as the brass core of LAGEOS II, are determined:
\begin{equation}
    \alpha_{C}=\sum_{\rm{all~A}} w_{CA}\alpha_A\;.
\end{equation}
The weightings $w_{CA}$ are the total mass fraction of each atomic species $A$ in material $C$. Through substitution of Eq.~\eqref{eq:Alpha_Definition} into the above expression, one may determine the effective dilaton charges for composite material $C$. Those relevant to the makeup of LAGEOS II and the Earth are given in Table~\ref{tab:dilaton_values}.

\subsection{Validity of the Linearised Force Approximation}\label{sec:liniarised_approx_breakdown}
\begin{figure}
    \centering
    \includegraphics[width=0.7\linewidth]{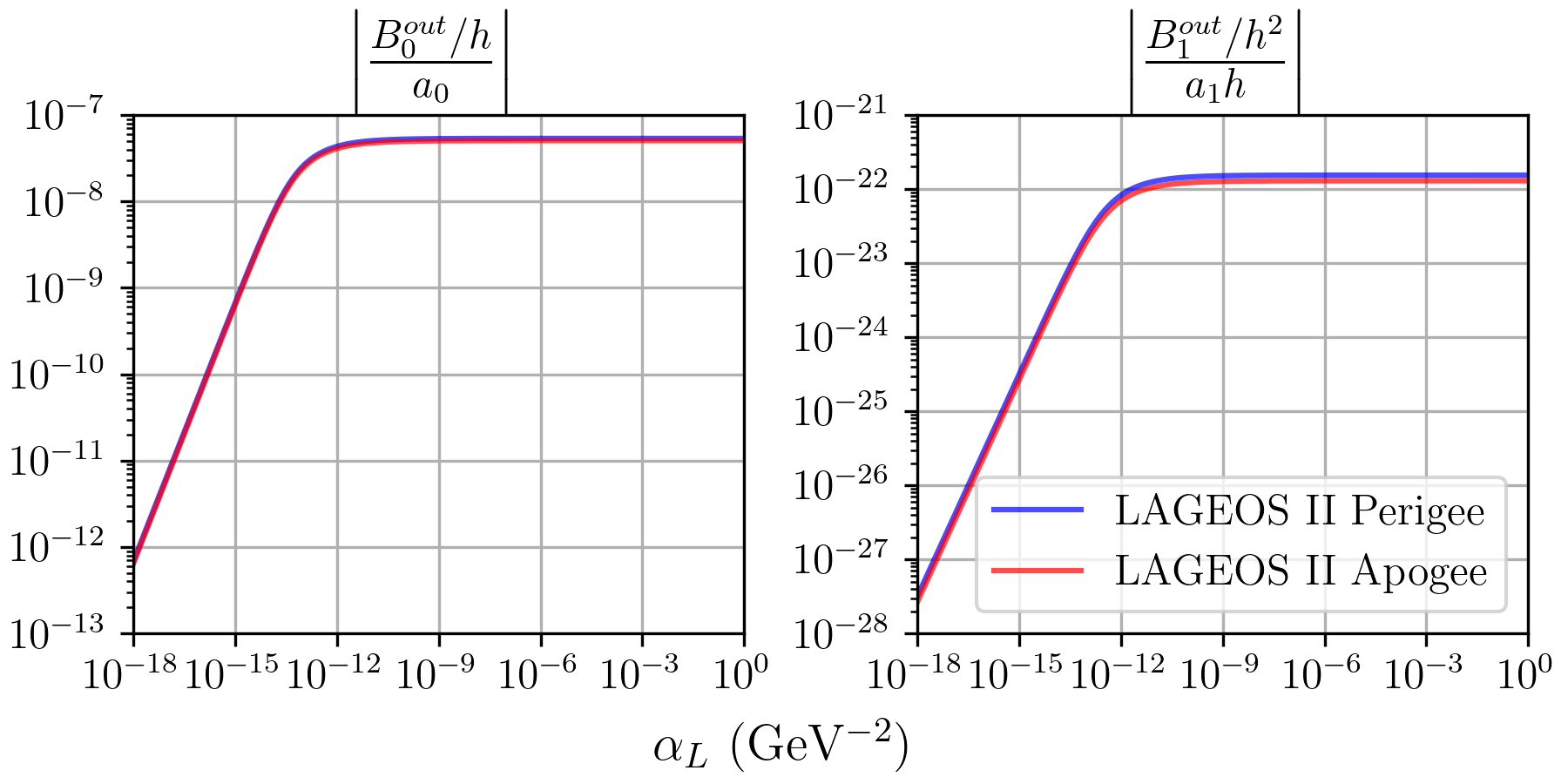}
    \caption{The quantities in Eq.~\eqref{eq:linerised_BC_limit_condtions}, plotted as a function of the coupling $\alpha_L$, which determine the validity of using linearised boundary conditions to describe the scalar field profile in the neighbourhood of the LAGEOS II satellite. For ease of visualisation, we have here used a composition-independent $\alpha$ such that $\alpha_L=\alpha_\oplus$. Whilst accounting for the composition dependence may subtly change the shape of the lines in the panels above, from the definitions of the coefficients $B^{out}_i$ given in Appendix~\ref{app:field_coeffs}, their upper bounds should not be changed. \textit{Red} corresponds to their value at the apogee of LAGEOS II's orbit, and \textit{blue} at the perigee. The highly circular orbit of LAGEOS II makes these difficult to distinguish. Values $\ll1$ are required for the field description to be valid at a given choice of $\alpha_L$. }
    \label{fig:linearisation validity}
\end{figure}
In Sec.~\ref{sec:solutions_near_earth}, our use of linearised boundary conditions requires that LAGEOS II should only perturb the scalar field background in a region $r_L\ll h$, where $h$ is the height of the satellite above the Earth. We will explicitly demonstrate that this requirement is satisfied.

For perturbations to the field profile from LAGEOS II to be sufficiently contained, then terms associated with the coefficients: $B_{L,i}^{out}$ in Eq.~\eqref{eq:linearised_solution} should be dominated by the terms originating in boundary conditions, associated with the coefficients $a_i$, before reaching $r_L\sim h$. This can be represented in the following inequalities:
\begin{equation}
    \left|\frac{B^{out}_{L,0}/h}{a_0}\right|\ll1\;, ~~~~~~~~~~~~~~~\left|\frac{B^{out}_{L,1}/h^2}{a_1h}\right|\ll1\;,
    \label{eq:linerised_BC_limit_condtions}
\end{equation}
which must hold for all heights $h=h(t)$ of the satellite above the Earth's surface throughout its trajectory. We plot the quantities in Eq.~\eqref{eq:linerised_BC_limit_condtions} in Fig.~\ref{fig:linearisation validity} for $h$ at both the perigee and apogee of LAGEOS II's orbit. We observe that the inequalities in Eq.~\eqref{eq:linerised_BC_limit_condtions} are well observed.

\subsection{Earth's Internal Structure}
\begin{figure}
    \centering
    \subfloat[]{\hspace{-2cm}\includegraphics[width=0.5\linewidth]{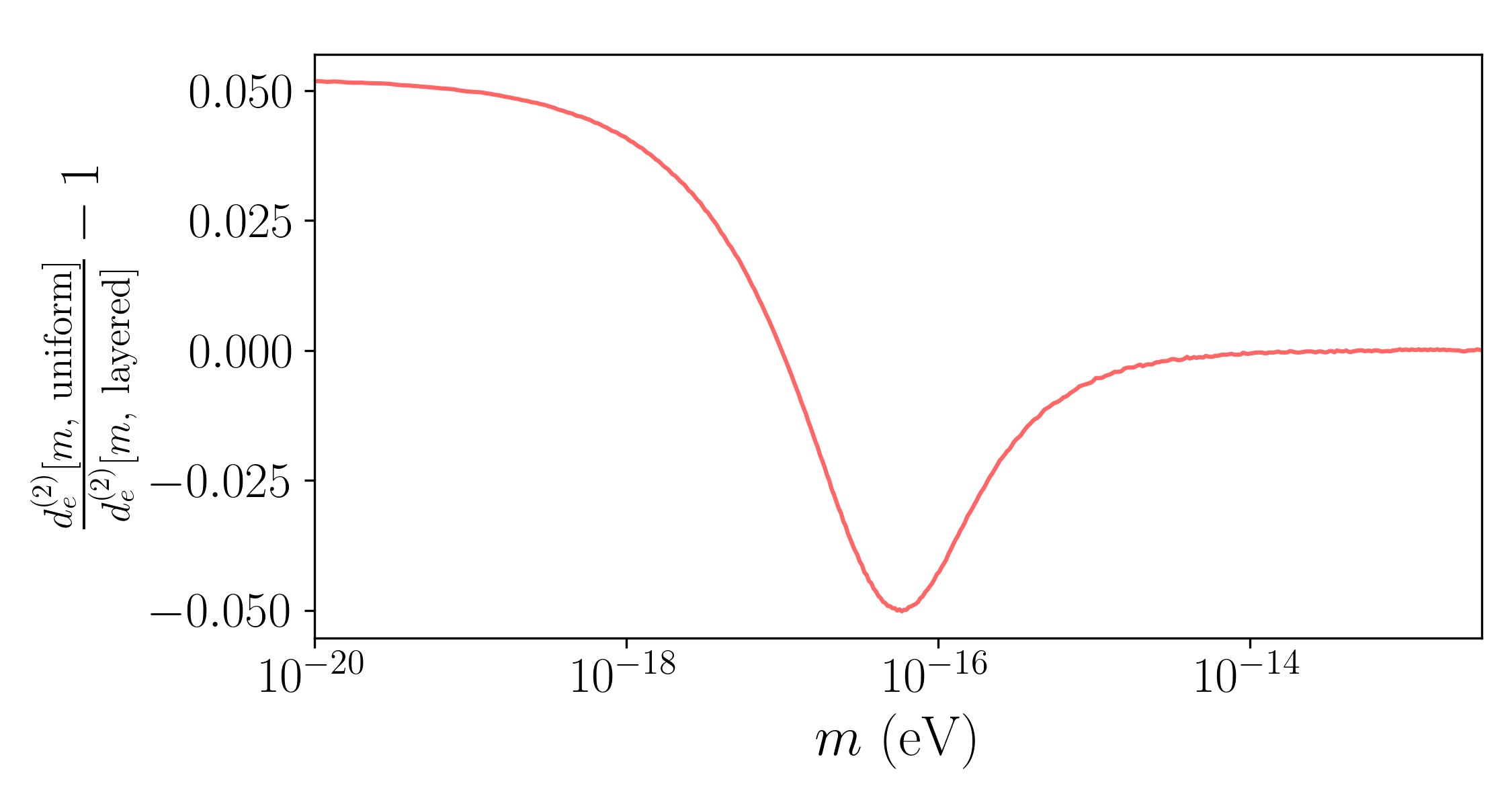}\label{uniform_layered_deviation_a}}\par
    \subfloat[]{\hspace{-2cm}\includegraphics[width=0.5625\linewidth]{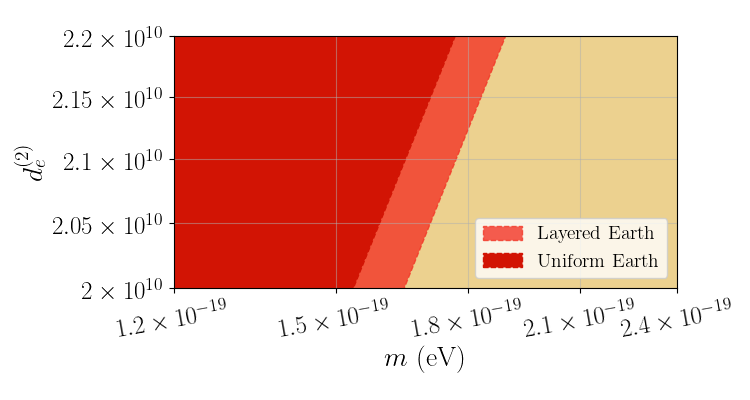}\label{uniform_layered_deviation_b}}%from 
    \caption{The effect of accounting for the Earth's non-trivial density structure - here exemplified for constraints on $d_e^{(2)}$. In panel (a) - the deviation of the boundary of the excluded region from pericentre precession in Fig.~\ref{subfig:a}, expressed as a function: $d_e^{(2)}(m)$, between a layered `iron-silicate' Earth description (see properties in Table~\ref{tab:spheres_properties}) and a uniform `iron' Earth, with constant density $5510~\rm{kg~m^{-3}}$. In panel (b) - the effect of choosing each Earth description on the constraints shown in Fig.~\ref{subfig:a} from LAGEOS II data. This plot has been significantly zoomed as to demonstrate this effect. Both ultimately demonstrate that the choice of Earth's internal structure has a small impact, so long as Earth's broader properties are accurately described.}
    \label{fig:uniform_layered_deviation}
\end{figure}

Whilst we aim to give a reasonable description of the Earth in our analysis, the layered description we use is still simplified compared to the planet's actual density and composition. We here query the effect of Earth's internal structure on the region of parameter space excluded through LAGEOS II's pericentre precession. We do so by comparing our excluded region with that generated in the case that Earth is a uniform sphere - density $5110~\rm{kg~m^{-3}}$. This is simply done by replacing $B_\oplus^{out}$ in Eq.~\eqref{eq:pericentre_precession} with the equivalent term from Eq.~\eqref{eq:spherical_solution}. As an example, in Fig.~\ref{fig:uniform_layered_deviation} we show the deviation in the boundary of the excluded region $d_e^{(2)}(m)$ for the constraints displayed in Fig.~\ref{subfig:a}, between the uniform sphere and layered sphere models. As anticipated, the bounds converge as $m$ increases. This owes to the fact that $d_e^{(2)}(m)$ for both the layered and uniform case increases monotonically, and that the values of $B^{out}_\oplus$ for each model tend to the same limit at large $d_e^{(2)}$, as we noted in Sec.~\ref{sec:Layered Sphere}. At smaller masses, we observe a $\sim(5\%)$ deviation between these models. As can be seen from Fig.~\ref{uniform_layered_deviation_b}, this ultimately has a negligible effect on the form of the bounds presented in Fig.~\ref{fig:Precession Constraints}. 

\subsection{Earth's Oblateness}\label{sec:spheroid}
It is noted in Ref.~\cite{Lucchesi:2014uza} that the largest corrections to the Keplerian description of the
orbital motion of LAGEOS II derive from Newtonian corrections to the Earth's gravitational field, as a consequence of its oblateness. Indeed, previous investigation has demonstrated that departure from sphericity in massive bodies can have a significant impact on screening effects in similar scalar models \cite{Burrage:2014daa}. We here compute the scalar profile around the Earth, approximated as a uniform oblate spheroid under our quadratic model of scalar interactions, and quantify its deviation from a similarly uniform, spherical description. Throughout this section, since we will be describing a system containing only the Earth, we will choose to work in terms of the coupling parameter $\alpha_\oplus$ between the field and Earth, rather than the individual dilaton coefficients $d_i^{(2)}$.   

\subsubsection{Earth as an oblate spheroid}
Considering a polar co-ordinate system centred on the Earth, we define its surface:
\begin{equation}
    R_{\oplus}(\theta)=\frac{a_{\oplus}\sqrt{1-e_{\oplus}^2}}{\sqrt{(1-e_{\oplus}^2)+e_{\oplus}^2\cos^2\theta}}\;,
    \label{eq:spheroid_surface}
\end{equation}
equivalent to tracing an ellipse with semi-major axis $a_{\oplus}$, eccentricity $e_{\oplus}$, and semi-minor axis:
\begin{equation}
    b_{\oplus}=a_{\oplus}\sqrt{1-e_{\oplus}^2}\;.
\end{equation}
We parameterise the Earth as having an eccentricity: $e_\oplus=0.0818$ \cite{WGS84}.

\subsubsection{Approximating Solutions for Small Eccentricities}
Splitting $\phi(\vec{r},t)=\phi_\infty\cos(mt)\varphi(r,\theta)$, field solutions to Eq.~\ref{eq:scalar_eom} must obey:
\begin{equation}
    \nabla^2\varphi=\alpha_\oplus\rho(r,\theta)\varphi(r,\theta)
\end{equation}
where $ \rho(r,\theta)=\rho_{\oplus}\Theta(R_{\oplus}(\theta)-r)$ defines the density profile for the oblate Earth. This can be separated into an interior and exterior solution:
\begin{equation}
    \varphi(r,\theta)=\begin{dcases}
        \sum_nA_n^{Ob}i_n(k_{\oplus}r)P_n(\cos\theta) & r<R_\oplus(\theta)\\
        1+\sum_n\frac{B_n^{Ob}}{r^{n+1}}P_n(\cos\theta)& r>R_\oplus(\theta)\;.
    \end{dcases}
\end{equation}
 $P_n$ are Legendre polynomials, and $i_n$ are modified spherical Bessel functions - both of the first kind. $k_\oplus=\sqrt{\alpha_\oplus\rho_\oplus}$ again gives the growth rate of the field inside matter. $A^{Ob}_n,B^{Ob}_n$ are constants of integration to be determined. Doing so requires imposing conditions of differentiability and continuity along the surface of the spheroidal Earth: $R_\oplus(\theta)$ - a condition which proves challenging to implement analytically due to the loss of spherical symmetry. We employ a semi-analytical method of evaluating boundary conditions, making use of the smallness of the Earth's eccentricity $e_\oplus$. Details of this method can be found in Appendix~\ref{app:oblate_solutions}, and yield a set of numerically determined values for the coefficients $A^{Ob}_n/B_n^{Ob}$, incorporating contributions up to $\mathcal{O}[e^2]$. 
 
\subsubsection{The Impact of Earth's Non-Sphericity}
\begin{figure}
    \centering
    \includegraphics[width=0.7\linewidth]{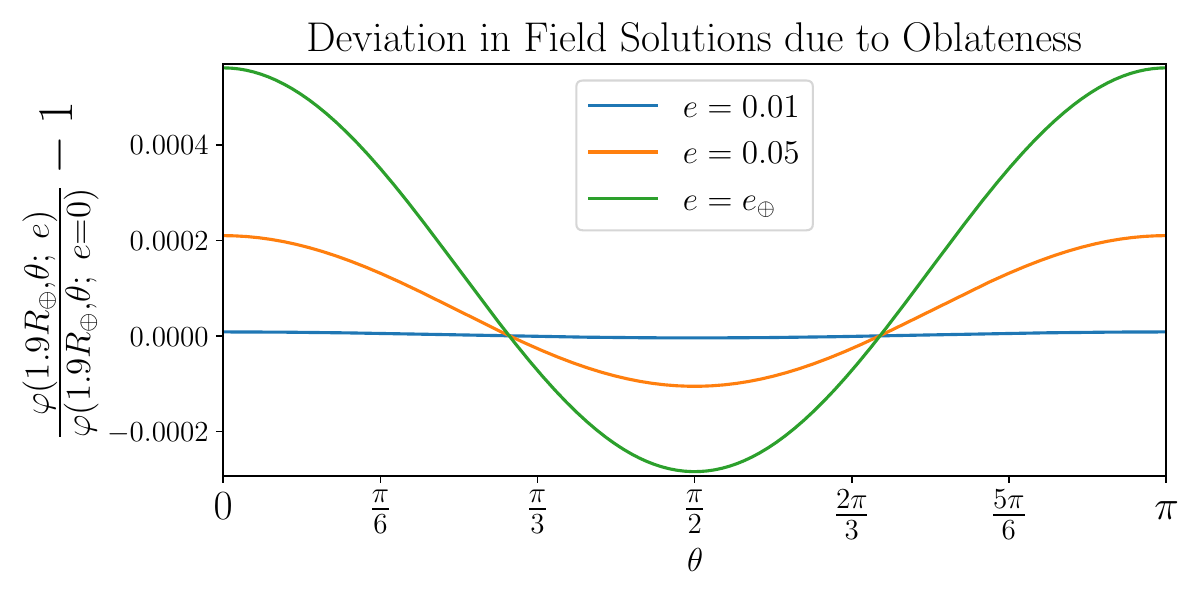}
    \caption{$\epsilon(\theta)$ is computed, comparing the angular field profile at the approximate orbital position of LAGEOS II around uniform spheroids of varying eccentricity. All spheroids have identical mass. The case $e=e_\oplus$ is shown in green. We observe that the total deviation between the spherical and oblate spheroid profiles, for the case $e=e_\oplus$ never exceeds $\sim10^{-3}$, providing reassurance that the oblateness of the Earth provides only small corrections to the previously calculated fifth forces, and pericentre precession. The coupling $\alpha_\oplus=10^{-26} ~\rm{GeV}^{-2}$ has been used here. As discussed in Sec.~\ref{sec:uniform_sphere}, this corresponds to a regime in which the field is more strongly screened in the vicinity of the Earth. See Fig.~\ref{fig:Scalar_Dev_Coupling} for an illustration of how $\epsilon(\theta)$ varies with the coupling strength $\alpha$.}
    \label{fig:Scalar_Dev}
\end{figure}
\begin{figure}
    \centering
    \includegraphics[width=0.7\linewidth]{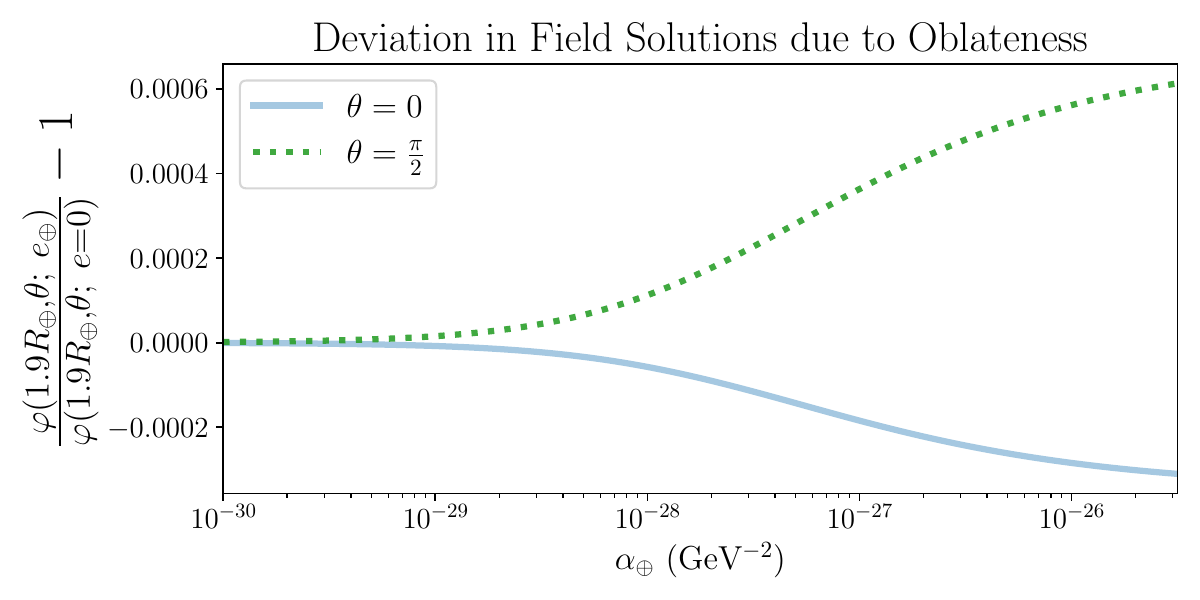}
    \caption{Deviation of the field profile around a spheroidal Earth, $e=e_\oplus$, from the profile around a spherical Earth $e=0$. Deviation is calculated at the orbital altitude of LAGEOS II: $r_L=1.9R_\oplus$, varying the coupling strength $\alpha_\oplus$. We plot up to an upper bound on $\alpha_\oplus$: $k_\oplus(a_\oplus-b_\oplus)\sim0.1$, above which our method breaks down. As we enter the strong coupling limit between the field and the Earth, $k_\oplus R_\oplus>1$, the correction to the equator and polar field values is always $<10^{-3}$.}
    \label{fig:Scalar_Dev_Coupling}
\end{figure}
To determine the deviation between the oblate and spherical solutions, we consider the quantity:
\begin{equation}
    \epsilon(\theta)=\frac{\varphi(r=1.9R_\oplus,\theta;e=e_\oplus)}{\varphi(r=1.9R_\oplus,\theta;e=0)}-1
\end{equation}
Where $1.9R_\oplus$ is the approximate orbital altitude of the LAGEOS II satellite, $e=e_\oplus$ indicates the scalar profile about an oblate spheroid Earth, and $e=0$ indicates the profile about a spherical Earth. Values $|\epsilon|\ll1$ are desired for the approximation that the Earth is spherical, not spheroidal, to be valid in our calculations. 

Fig.~\ref{fig:Scalar_Dev} illustrates angular variation in $\epsilon$. As expected, we observe that the scalar profiles increasingly deviate from the spherical solution with increasing spheroid eccentricity, and that the field is increasingly screened near the Earth's equator, where a point at fixed distance from the Earth's centre is most proximal to matter. Fig.~\ref{fig:Scalar_Dev_Coupling} illustrates how $\epsilon$ varies with the coupling strength $\alpha_\oplus$ at two fixed points in space: one above a pole of the Earth, and the other above its equator. Both show Earth's oblateness has an $|\epsilon|< \mathcal{O}[10^{-3}]$ impact on the scalar profile experienced by the satellite. We note that an upper bound on the range of $\alpha_\oplus$ considered is set by a caveat with our method - described in Appendix~\ref{app:oblate_solutions} - which is accurate only for $k_\oplus\Delta R_\oplus\sim k(a_\oplus-b_\oplus)\ll1$. This sets a maximum: $\alpha_\oplus\ll3.7\times10^{-25}~\rm{GeV}^{-2}$ or $d_e^{(2)},~d_{\hat{m}}^{(2)}-d_g^{(2)},~d_{m_e}^{(2)}-d_g^{(2)} \lesssim10^{15}$ in the language of Fig.~\ref{fig:Precession Constraints}. Whilst not ideal, prior understanding that the field profiles will rapidly converge to some maximally screened configuration at large $\alpha_\oplus>\frac{1}{\rho_\oplus R_\oplus^2}$, and observation of the flattening of the curves demonstrated in Fig.~\ref{fig:Scalar_Dev_Coupling} with increasing $\alpha_\oplus$ mean that deviation beyond $|\epsilon|\sim\mathcal{O}[10^{-3}]$ should not be expected. 

We here conclude that, because corrections from Earth's oblateness would constitute small corrections to fifth force effects, which already necessarily contribute small corrections to the motion LAGEOS II, it is indeed safe to neglect them in our calculations.  
\subsection{Dark Matter Wind}
\label{sec:DM_wind}

We consider here the possible influence of an incoherent dark matter wind on the constraints discussed in Sec.~\eqref{sec:constraints}. This requires that the boundary conditions describing the field away from matter in Eq.~\eqref{eq:homogeneous} are modified to incorporate a distribution of momenta with some net direction. To do so, we focus on the calculation of $\langle\nabla\phi^2\rangle$, to which fifth forces mediated by quadratically coupled scalars are sensitive. We evaluate this quantity for two key sets of field profiles: $\phi_w$ (defined in this section) describing the field amplitude around the Earth in the presence of the wind, and $\phi_S$ (defined in Eq.~\eqref{eq:spherical_solution}), describing the field profile around the Earth in its absence. We are specifically interested in these fields evaluated at the orbital radius of LAGEOS II $\sim1.9R_\oplus$. We note that for simplicity in these calculations, we are again approximating the Earth as a sphere of uniform density.
Whilst $\langle\nabla\phi^2_S\rangle$ may be trivially evaluated, this is not so for $\langle\nabla\phi^2_w\rangle$ due to the breaking of spherical symmetry introduced by the dark matter wind. In its calculation, we follow from Refs.~\cite{Hui:2020hbq, Brzeminski:2026rox, VanTilburg:2024xib, Gan:2025nlu}. We model  the dark matter wind as an incoherent ensemble of field modes:
\begin{equation}
    \phi_w=\phi_\infty\int d^3\vec{p} \sqrt{f(\vec{p};\vec{p}_0)}~\RE\left[{\varphi}(\vec{r},\vec{p})e^{i\left(\omega(p)t+\chi_{\vec{p}}\right)}\right]\;.
    \label{eq:wind_int_over_modes}
\end{equation}
where the function $f(\vec{p};\vec{p}_0)$ describes the distribution of momenta, which for illustrative purposes we model using boosted Maxwell-Boltzmann statistics:
\begin{equation}
    f(\vec{p};\vec{p}_0) = \left(\frac{1}{2\pi\sigma_p^2}\right)^{3/2}e^{-\frac{(\vec{p}-\vec{p}_0)^2}{2\sigma^2}}\;.
    \label{eq:wind_momentum_dist}
\end{equation}
Here, $p_0$ is principally determined from the velocity of the Earth, and the circular velocity of the solar system, and $\sigma$ from the velocity dispersion of the Dark Matter Halo. Given the similar magnitudes of these velocities, we approximate $\beta\approx\tfrac{p_0}{m}\approx\tfrac{\sqrt{2}\sigma}{m} \sim 10^{-3}$ \cite{Drukier:1986tm}. The spatial component of the field profile ${\varphi_w}$ satisfies:
\begin{equation}
    \nabla^2{\varphi_w}+(p^2-\rho\alpha){\varphi_w}=0 \;.
\end{equation}
Implementing boundary conditions: ${\varphi_w}(r\rightarrow\infty,\theta;\vec{p}) = \phi_\infty e^{-i\vec{p}\cdot \vec{r}}$ around a uniform-density Earth yields solutions of the form:
\begin{equation}
    \varphi_w(\vec{r};\vec{p})=\begin{dcases}
        \sum_{l=0}^\infty(-i)^l(2l+1)(j_l(pr)+A^{out}_l(p)~h^{(2)}_l(pr))P_l(\cos\theta) &r>R_\oplus\\
        \sum_{l=0}^\infty(-i)^l(2l+1)A^{in}_l(p)j_l(qr)P_l(\cos\theta) &r<R_\oplus
    \end{dcases}\;,
    \label{eq:Wind_Legendre_Modes}
\end{equation}
where for $\vec{p}=\vec{p}(p,\theta_p,\psi_p)$, $\vec{r}=\vec{r}(r,\theta_r,\psi_r)$ the angle separating these vectors $\hat{p}\cdot\hat{r}=\cos\theta = \sin\theta_r\cos\psi_r\sin\theta_p\cos\psi_p  + \sin\theta_r\sin\psi_r\sin\theta_p\sin\psi_p + \cos\theta_r\cos\theta_p$. Recognising the cylindrical symmetry of the system, this may be simplified by setting $\psi_r=0$. We henceforth discard the internal solutions ($r<R_\oplus$) as these are not needed for the motion of LAGEOS II. For the external solutions ($r>R_\oplus$), the sum over spherical Bessel functions of the first kind, $j_l$, gives the contribution to the profile from the incoming dark matter wind, which may be observed through the identity:
\begin{equation}
    e^{-i\vec{p}\cdot\vec{r}}=\sum_{l=0}^\infty (-i)^l(2l+1)j_l(pr)P_l(\cos\theta)\;.
    \label{eq:exp_identity}
\end{equation}
Similarly, the sum over spherical Hankel functions $h_l^{(2)}$ gives the field contribution from waves scattering on the Earth. The coefficient $A_l^{out}$ is defined:
\begin{equation}
            A_l^{out}= -\frac{p j_{l+1}(p R) j_l(q R)-q~ j_l(p R) j_{l+1}(q R)}{p j_l(q R) h^{(2)}_{l+1}(p R)-q j_{l+1}(q R) h_l^{(2)}(p R)}
\end{equation}
where $q=\sqrt{p^2-\rho\alpha}$. Note that $q(p=0)=ik_S$.
Finally, $\chi_{\vec{p}}\in[0,2\pi]$ is a random phase, drawn uniformly. We express the incoherence of the incoming dark matter wind by defining the inner product: 
\begin{equation}
    \langle e^{i\chi_{\vec{p}}}e^{-i\chi_{\vec{p}'}}\rangle=\delta^{(3)}(\vec{p}-\vec{p}')\;.
    \label{eq:wind_phase_avg}
\end{equation} 
When applied as in: $\langle\nabla\phi^2\rangle$, this inner product serves to average $\nabla\phi^2$ over the values of the phases $\chi_{\vec{p}}$, and ultimately has the effect of averaging out time-oscillatory behaviour. For this average to well describe the properties of the field over a time period $\tau$, then it is required that $\tau\gg\tau_c$, where $\tau_c\approx\frac{1}{\beta^2m}$ is the coherence time of the field. Choosing $\tau\sim 13~\rm years$, the period over which the datum in Eq.~\eqref{eq:anomalous_precession} is produced \cite{Lucchesi:2014uza}, then this sets a lower limit on the field mass $m>10^{-18}~\rm eV$.

With the above principles established, we may compute the value of $\langle\nabla\phi^2_w\rangle$ at the orbital radius of LAGEOS II $r\simeq1.9R_\oplus$. In particular, we focus on calculating the radial component of the gradients: $\langle\partial_r\phi^2_i\rangle$ (where $i\in w,S$) since it is the only non-zero component of $\langle\nabla\phi^2_S\rangle$, and hence is most useful for comparison. 

\begin{figure}
    \centering
    \includegraphics[width=0.6\linewidth]{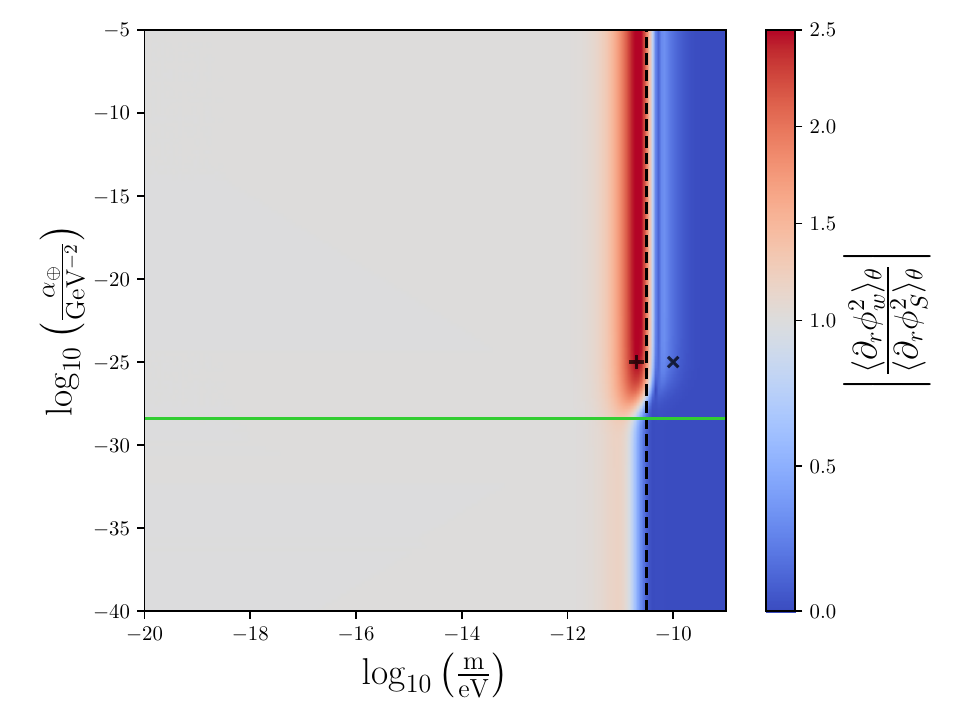}
    \caption{The ratio of the average value of the field gradient in the presence of a dark matter wind, with field momenta distributed as in Eq.~\eqref{eq:wind_momentum_dist}, with $p_0=10^{-3}m$, over a range of field masses $m$ and coupling couplings to matter $\alpha$. The \textit{black dashed line} indicates the threshold $p_0R_\oplus=1$, where the wavelength of the field becomes comparable to the Earth's radius. The \textit{green solid line} indicates the threshold $\alpha_\oplus\rho_\oplus R_\oplus^2=1$, where the skin depth of the field inside the Earth becomes comparable to its radius. Grey regions indicate where scalar-mediated fifth forces in the presence of a dark matter wind are similar in magnitude, on average, to those in the absence of the wind. Red and blue regions show where the average force experienced by LAGEOS II might be enhanced or suppressed by the dark matter wind respectively. }
    \label{fig:wind}
\end{figure}

For ease of presentation, we additionally choose to average the $\langle\partial_r\phi^2_i\rangle$ over the angular co-ordinate $\theta_r$, which we denote $\langle\partial_r\phi^2_i\rangle_\theta$. This is equivalent to determining the average force experienced by LAGEOS II along a circular orbit, where the vector $\vec{p}_0$ runs parallel to the plane of said orbit. On such a trajectory, the satellite will experience the most variation in the scalar field profile due to the presence of the dark matter wind. Following this, we are ultimately led to evaluate the integral: 
\begin{equation}
    \langle\partial_r\phi^2_w\rangle_{\theta} = 2\pi\phi_\infty^2 \left(\frac{1}{\pi p_0^2}\right)^{3/2}\sum_{l}\frac{1}{2l+1}\int_0^\infty dp ~p^2~\left[\partial_rF_l(p,1.9 R_\oplus)F^*_{l}(p,1.9 R_\oplus)+{\rm{cc.}}\right]i_0\left(\frac{2p}{p_0}\right)e^{-\frac{p^2}{p^2_0}-1}
    \label{eq:wind_field_gradient}
\end{equation}
Where $F_l$ gives the radial component of the $l^{\rm{th}}$ harmonic mode in Eq.~\eqref{eq:Wind_Legendre_Modes}, and $i_0$ is the $0^{\rm{th}}$ order modified spherical Bessel function of the first kind. This is too complex to evaluate analytically and must be approached numerically. Further detail as to the derivation of this expression, and its computation is provided in Appendix~\ref{app:DM_wind}. Thus, we demonstrate the effect of a dark matter wind on fifth forces in Fig.~\ref{fig:wind} through the ratio: $\left|\langle\partial_r\phi_w^2\rangle_\theta/\langle\partial_r\phi_S^2\rangle_\theta\right|$, across a range of field masses and couplings. 

\begin{figure}
    \centering
    \subfloat[]{\includegraphics[width=0.5\linewidth]{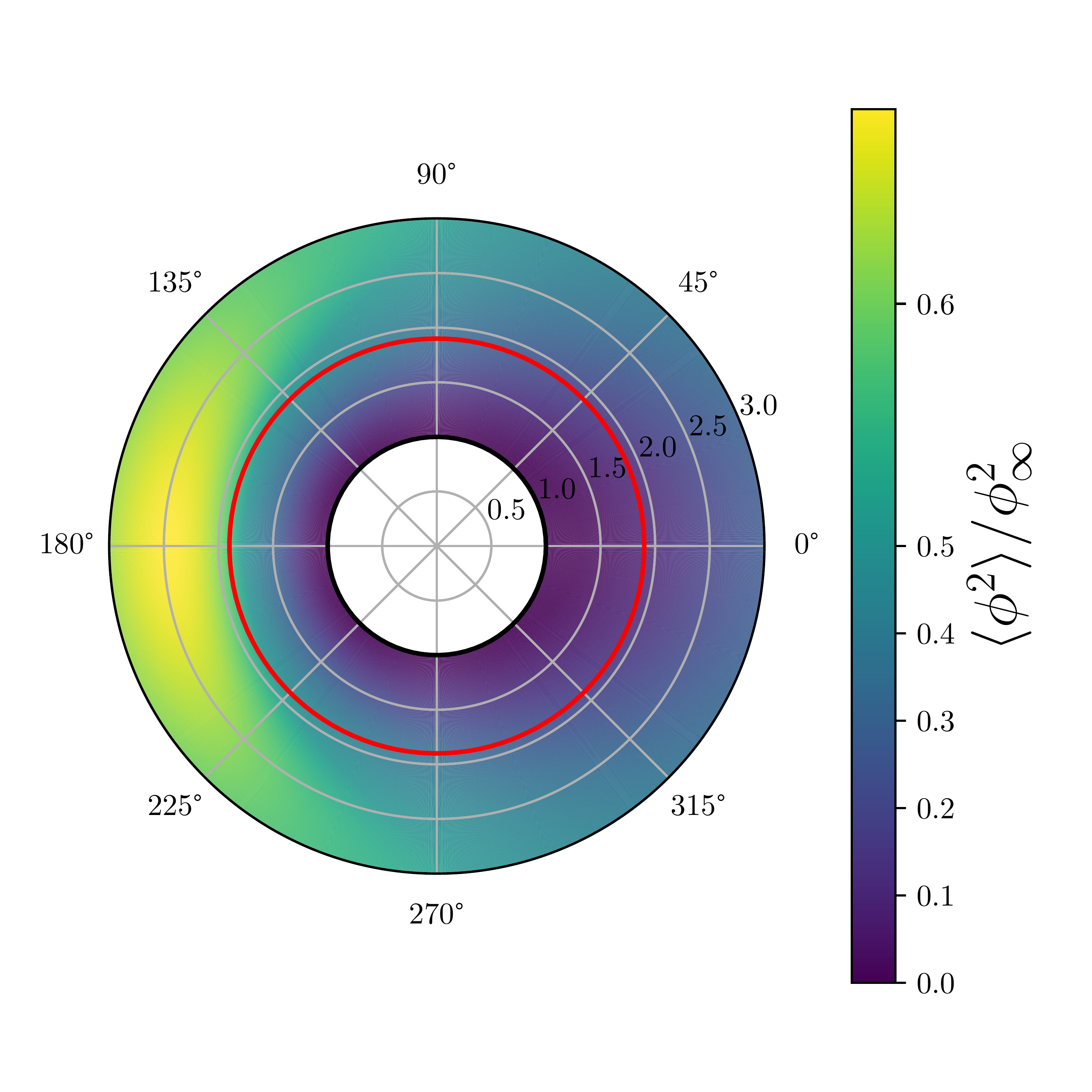}\label{}}\hfill
    \subfloat[]{\includegraphics[width=0.5\linewidth]{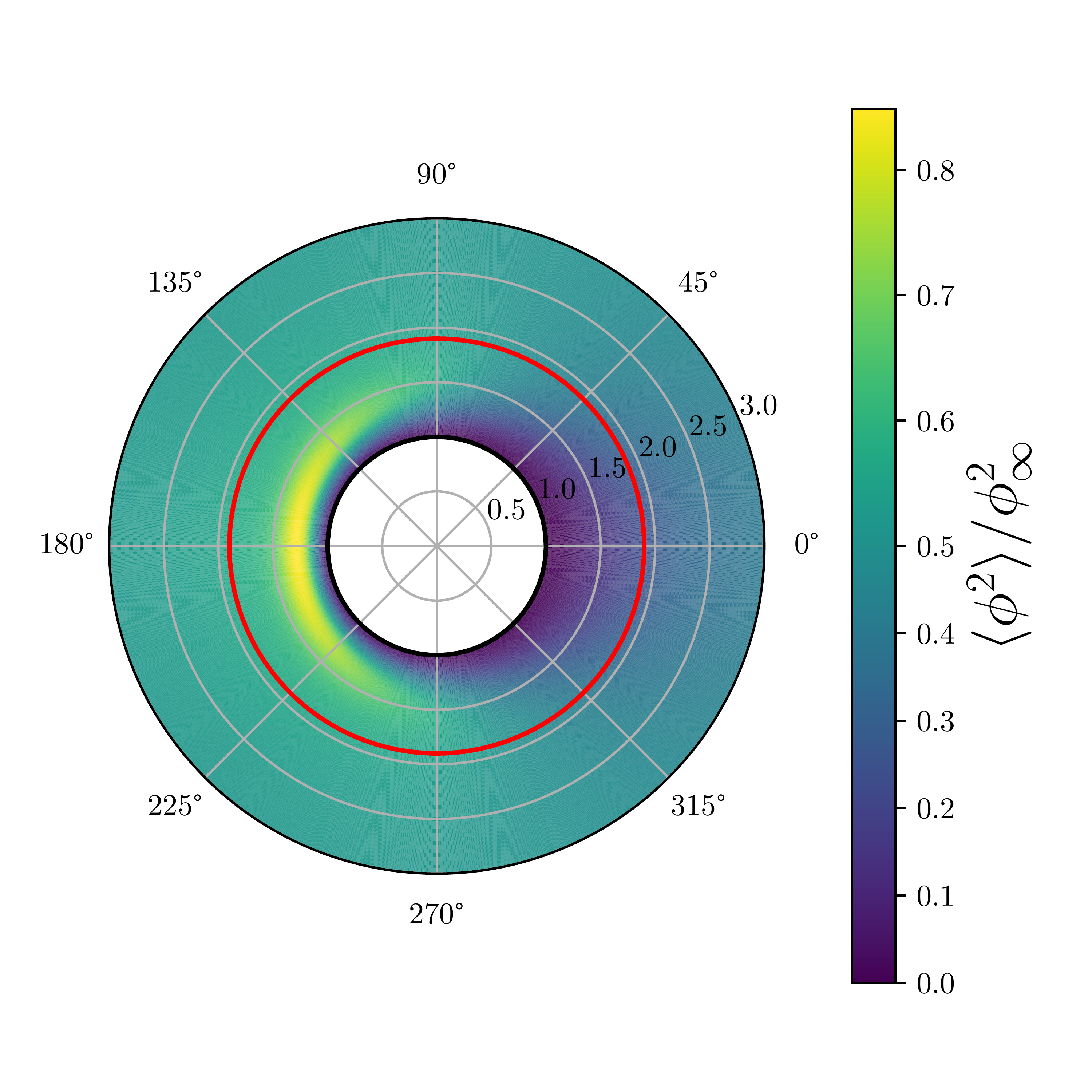}\label{}}\hfill
     \caption{The ensemble-averaged field-squared profile around the Earth - excluding the field profile within the Earth - for field masses $m=2\times10^{-11}\rm{~eV}$ (panel (a) ), and $m=10^{-10}\rm{~eV}$ (panel (b) ). The radial co-ordinate here is given in units of $R_\oplus$, with the \textit{black solid line} indicating the surface of the Earth, and the \textit{red solid line} indicating the approximate orbital position of LAGEOS II. Both profiles assume a coupling to Earth $\alpha=10^{-25}~\rm{GeV^{-2}}$. These parameter choices are illustrated on Fig.~\ref{fig:wind} by the ``$+$" and ``$\times$" markers respectively.}
    \label{fig:wind_profiles}
\end{figure}

For $p_0R_\oplus\ll1$, we observe that the field profiles in the presence and absence of a dark matter wind are in good agreement. However, where $p_0R_\oplus\gtrsim1$, we observe a significant deviation between the cases. At $p_0R_\oplus\sim1$, the value of $\langle\partial_r\phi^2_w\rangle_\theta$ is enhanced above the homogeneous case. In Fig.~\ref{fig:wind_profiles}, where we demonstrate $\langle\phi^2_w\rangle$ itself, we observe that this derives from the constructive interference of a ``bow-shock" type effect. At larger $m$, which correspond to larger $p_0$, this `bow shock' is shifted closer to the surface of the Earth. At radii beyond this `bow shock', Fig.~\ref{fig:wind_profiles} demonstrates a `washing out' of field gradients. Hence, for sufficiently massive fields, where the bow shock is closer to the Earth than LAGEOS II, $\langle\partial_r\phi^2_w\rangle_\theta$ appears suppressed, leading to the blue region in Fig.~\ref{fig:wind}. 

In principle, we could attempt to rescale our constraints in Fig.~\ref{fig:Precession Constraints}, using the average suppression/enhancement of forces, induced by the dark matter wind, shown in Fig.~\ref{fig:wind}. However, since this suppression/enhancement is not spherically symmetric about the Earth, its impact on the orbit of LAGEOS II, and subsequently on its apparent pericentre precession, would be non-trivial. Hence, we choose to display $p_0R_\oplus\sim1$ in Fig.~\ref{fig:Precession Constraints} as a simple threshold, indicating where constraints become less trustworthy, and leave more complete rescaling of constraints on parameter space $p_0R_\oplus\gtrsim1$ as an avenue for future work. 

\section{Conclusions}
Ultralight scalars are a possible means by which to explain the apparent dark matter content of the universe, and certain models are anticipated to feature quadratic leading-order couplings to Standard Model fields. The phenomenology of quadratically coupled scalars is such that, in the vicinity of matter, their amplitude may be suppressed. In a previous work, we showed that this could lead to a relaxation of existing constraints from tabletop and satellite experiments at strong couplings. In this work, we have demonstrated that fifth forces on satellites around the Earth can yield pericentre precession of their obits. Using the anomalous pericentre precession of the LAGEOS II satellite, we have constrained the mass, and coupling strength of the field to different light Standard Model fields. We confirm that these constraints are able to cover parameter space at strong couplings, where the aforementioned relaxation is required. At weaker couplings between matter and the scalar field, we find these constraints to be competitive with existing constraints from satellite and tabletop level experiments. Future work yielding more precise measurements of the anomalous pericentre precession of satellites, possibly through a greater abundance of data, more precise laser ranging methods or improved descriptions of the Earth's gravity, may prove an effective means to further constrain quadratically coupled ultralight dark matter.

\section*{Acknowledgements}
We would like to thank Aurelien Hees for useful discussions in the early stages of this project.  CB and ET were supported in part by STFC Consolidated
Grant [Grant No. ST/T000732/1]. AM is supported by EPSRC Quantum Technologies Doctoral
Training Partnership (DTP) 2024-25. 
ET is supported by the European Union’s Horizon 2020
research and innovation programme under the Marie Skłodowska-Curie grant
agreement No 101204903 (APARAX).
For open access purposes, the authors have applied a CC BY public
copyright license to any author accepted manuscript version arising from this submission.

\section*{Data Availability Statement}
In case of legitimate interest, the authors will provide the Python code used for producing the
plots in this work on request.

\bibliography{sample}
\appendix 

\section{Linearised Field Coefficients} \label{app:field_coeffs}
Coefficients for the scalar field profile around a layered spherical body, as discussed in Sec.~\ref{sec:Layered Sphere}, and Sec.~\ref{sec:solutions_near_earth}. Specifically, the coefficients below are those for the case where the scalar field background is linear, as in Eq.~\eqref{eq:linearised_solution}. Coefficients for field solutions on a flat background such as in Eq.~\eqref{eq:layered_sphere_solutions}, needed to describe the scalar profile about the Earth, are determined by setting $a_0=1$ and $a_1=0$.

\begin{align}
    A_{0}^{int}=&\frac{a_0 k^i}{k^i \cosh \left(k^i R^i\right) \cosh \left(k^e \left(R^e-R^i\right)\right)+k^e \sinh \left(k^i R^i\right) \sinh \left(k^e \left(R^e-R^i\right)\right)}
    \\
    \notag
    \\
    A_{0}^{ext}=&\frac{a_0 \left(k^{i} \cosh \left(k^{i} R^i\right) \cosh \left(k^e R^i\right)-k^e \sinh \left(k^i R^i\right) \sinh \left(k^e R^i\right)\right)}{k^i \cosh \left(k^i R^i\right) \cosh \left(k^e \left(R^e-R^i\right)\right)+k^e \sinh \left(k^i R^i\right) \sinh \left(k^e \left(R^e-R^i\right)\right)}
    \\
    \notag\\
    B^{ext}_{0} =& \frac{a_0 \left(k^e \sinh \left(k^i R^i\right) \cosh \left(k^e R^i\right)-k^i \sinh \left(k^e R^i\right) \cosh \left(k^i R^i\right)\right)}{k^i \cosh \left(k^i R^i\right) \cosh \left(k^e \left(R^e-R^i\right)\right)+k^e \sinh \left(k^i R^i\right) \sinh \left(k^e \left(R^e-R^i\right)\right)}
    \\
    \notag\\
    B_{0}^{out}=&-a_0 \left(R^e-\frac{k^i \tanh \left(k^e \left(R^e-R^i\right)\right) \coth \left(k^i R^i\right)+k^e}{{k^e}^2 \tanh \left(k^e \left(R^e-R^i\right)\right)+k^i k^e \coth \left(k^i R^i\right)}\right)
    \label{eq:Coefficient_B0_Out}
    \\
    \notag\\
    A_{1}^{int}=&\frac{-3 a_1 {k^i}^2 k^e R^i R^e}{\left(\splitdfrac{{k^i}^2 k^e R^i \sinh \left(k^i R^i\right) \cosh \left(k^e \left(R^i-R^e\right)\right)}{+\sinh \left(k^e \left(R^e-R^i\right)\right) \left(\left({k^i}^2-{k^e}^2\right) \sinh \left(k^i R^i\right)+k^i {k^e}^2 R^i \cosh \left(k^i R^i\right)\right)}\right)}
    \\
    \notag\\
    A_{1}^{ext} =&\frac{3 a_1 R^e\left(\splitdfrac{\left({k^e}^2-{k^i}^2\right) \sinh \left(k^i R^i\right) \cosh \left(k^e R^i\right)+}{k^i k^e R^i \left(k^i \sinh \left(k^i R^i\right) \sinh \left(k^e R^i\right)-k^e \cosh \left(k^i R^i\right) \cosh \left(k^e R^i\right)\right)}\right)}{\left(\splitdfrac{{k^i}^2 k^e R^i \sinh \left(k^i R^i\right) \cosh \left(k^e \left(R^i-R^e\right)\right)}{+\sinh \left(k^e \left(R^e-R^i\right)\right) \left(\left({k^i}^2-{k^e}^2\right) \sinh \left(k^i R^i\right)+k^i {k^e}^2 R^i \cosh \left(k^i R^i\right)\right)}\right)}
    \\\notag\\
B_{1}^{ext}=&\frac{3 a_1 R^e \left(\splitdfrac{\left({k^e}^2-{k^i}^2\right) \sinh \left(k^i R^i\right) \sinh \left(k^e R^i\right)}{+k^i k^e R^i \left(k^i \sinh \left(k^i R^i\right) \cosh \left(k^e R^i\right)-k^e \sinh \left(k^e R^i\right) \cosh \left(k^i R^i\right)\right)}\right)}{\left(\splitdfrac{{k^i}^2 k^e R^i \sinh \left(k^i R^i\right) \cosh \left(k^e \left(R^i-R^e\right)\right)}{+\sinh \left(k^e \left(R^e-R^i\right)\right) \left(\left({k^i}^2-{k^e}^2\right) \sinh \left(k^i R^i\right)+k^i {k^e}^2 R^i \cosh \left(k^i R^i\right)\right)}\right)}
\end{align}
\footnotesize
\begin{equation} 
B_{1}^{out} = \frac{a_1 R^e \left(\splitdfrac{-\sinh \left(k^e \left(R^e-R^i\right)\right) \left(\splitdfrac{\left({k^i}^2 \left({k^e}^2 R^e \left(R^e-3 R^i\right)+3\right)-{k^e}^2 \left({k^e}^2 {R^e}^2+3\right)\right) \sinh \left(k^i R^i\right)}{+k^i {k^e}^2 R^i \left({k^e}^2 {R^e}^2+3\right) \cosh \left(k^i R^i\right)}\right)}{+{k^e} \cosh \left({k^e} \left(R^e-R^i\right)\right) \left(\splitdfrac{-\left({k^i}^2 \left(R^i \left({k^e}^2 {R^e}^2+3\right)-3 {R^e}\right)+3 {k^e}^2 {R^e}\right) \sinh \left(k^i R^i\right)}{+3 k^i {k^e}^2 R^i R^e \cosh \left(k^i R^i\right)}\right)}\right)}{{k^e}^2 \left(\splitdfrac{{k^i}^2 k^e R^i \sinh \left(k^i R^i\right) \cosh \left(k^e \left(R^i-R^e\right)\right)}{+\sinh \left(k^e \left(R^e-R^i\right)\right) \left(\left({k^i}^2-{k^e}^2\right) \sinh \left(k^i R^i\right)+k^i {k^e}^2 R^i \cosh \left(k^i R^i\right)\right)}\right)}
\end{equation}
\normalsize
\\

\section{Oblate Solutions}
\label{app:oblate_solutions}
Continuity and differentiability of the field are required at the boundary:
\begin{equation}
    \varphi(r\rightarrow R_\oplus(\theta)^+,\theta)=\varphi(r\rightarrow R_\oplus(\theta)^-,\theta)
    \label{eq:continuity}
\end{equation}
\begin{equation}
    \nabla\varphi(r\rightarrow R_\oplus(\theta)^+,\theta)=\nabla\varphi(r\rightarrow R_\oplus(\theta)^-,\theta)
    \label{eq:differentiability}
\end{equation}
We approximate solutions to these equations through two stages. Foremost, we perform a Legendre series of Eq.~\eqref{eq:spheroid_surface}:
\begin{equation}
    \begin{split}R_\oplus(\theta)&\approx R_\oplus^{(0)}+R_\oplus^{(2)}P_2(\cos\theta)+\mathcal{O}[e^4]\\
    &=R_\oplus^{(0)}+\Delta R_\oplus(\theta)
\end{split}
\end{equation}
Whereby terms terms scale in orders of $e_{\oplus}$:
\begin{equation}    
    \begin{split}    
        R_\oplus^{(0)}&=a_{\oplus}\left(1-\frac{e^2}{6} + \mathcal{O}[e^4]\right)\\
        R_\oplus^{(2)}&=a_{\oplus}\left(-\frac{e^2}{3}+ \mathcal{O}[e^4]\right)\\
    \end{split}
\end{equation}
Second, we perform a linearisation of the boundary conditions in Eqs.~\eqref{eq:continuity}, \eqref{eq:differentiability}.
Assuming $k\Delta R_\oplus\ll1$, Eqs.~\eqref{eq:continuity} becomes:
\begin{equation}
\begin{split}
    &\sum_n\left(\left[A^{Ob}_ni_n(kR_\oplus^{(0)})-B_n^{Ob}\frac{1}{(R_\oplus^{(0)})^{n+1}}\right]P_n(\cos\theta)\right.\\&+\left.\left[A^{in}_nki'_n(kR_\oplus^{(0)})+B^{Ob}_n\frac{(n+1)}{(R_\oplus^{(0)})^{n+2}}\right]\Delta R_\oplus(\theta)P_n(\cos\theta)\right)=1
    \label{eq:approximation_continuity}
\end{split}
\end{equation}
and Eq.~\eqref{eq:differentiability} becomes:
\begin{equation}
    \begin{split}
    &\sum_n\left(\left[A^{Ob}_nki'_n(kR_\oplus^{(0)})-B^{Ob}_n\frac{-(n+1)}{(R_\oplus^{(0)})^{n+2}}\right]P_n(\cos\theta)\right.\\&+\left.\left[A^{Ob}_nk^2i''_n(kR_\oplus^{(0)})-B^{Ob}_n\frac{(n+1)(n+2)}{(R_\oplus^{(0)})^{n+3}} \right]\Delta R_\oplus(\theta)P_n(\cos\theta)\right)=0        
    \end{split}
    \label{eq:approximation_deriv_condition}
\end{equation}
Under this approximation, if at the surface the field is continuous, and differentiable in the radial direction, then its differentiability in the $\hat{\theta}$ direction is guaranteed. To construct a set of equations, with which to determine $A^{in/out}_n$, it is necessary to expand $P_n(\cos(\theta))\Delta R_\oplus(\theta)$ into a single series:
\begin{equation}
    P_n(\cos(\theta))\Delta R_\oplus(\theta)=\sum _pc_{pn}P_p(\cos(\theta))
\end{equation} 
This may be achieved through the use of the triple product identity:
\begin{equation}
    \frac{1}{2}\int dx P_m(x)P_n(x)P_p(x)=\left\{\begin{matrix}
        m & n & p\\0 & 0 & 0
    \end{matrix}\right\}^2
    \label{eq:wigner_triple}
\end{equation}
where the term in curved parentheses is the ``Wigner $3j$" symbol. Expecting that terms in $A_n^{Ob},~B_n^{Ob}$ scale in powers of $e$, we assume that the initial summation over $n$ contains terms up to $n=4$. Noting the symmetry of the system, we also drop odd $n$ terms. With these principles established, Eqs.~\eqref{eq:approximation_continuity}, \eqref{eq:approximation_deriv_condition} may be solved as a system of linear equations. We accomplish this numerically.

\section{Dark Matter Wind Effects: Computation}
\label{app:DM_wind}

We continue, following from the notation established up to Eq.~\eqref{eq:wind_field_gradient}. We wish to calculate the value of $\langle\partial_r\phi^2\rangle$ where angled brackets here represent an average over phases $\chi_{\vec{p}}$. Substituting the expression for $\varphi$ in Eq.~\eqref{eq:Wind_Legendre_Modes} into Eq.~\eqref{eq:wind_int_over_modes}, this takes the form:
\begin{equation}
    \begin{split}
    \langle\partial_r\phi^2\rangle = \langle\partial_r\int d^3p\int d^3p'\sqrt{f(\vec{p};\vec{p}_0)f(\vec{p}';\vec{p}_0)}\sum_{l~l'}
    &\left[\frac{1}{2}\left(F_l(r,p)P_l(r)e^{i(\omega_pt+\chi_{\vec{p}})}+{\rm{cc.}}\right)\right.\\&\times\left.\frac{1}{2}\left(F_{l'}(r,p')P_{l'}(\cos\theta)e^{i(\omega_{p'}t+\chi_{\vec{p}'})}+{\rm{cc.}}\right)\right]\rangle    
    \end{split}
\end{equation}
which, through inserting and evaluating Eq.~\eqref{eq:wind_phase_avg} is simplified:
\begin{equation}
    \langle\partial_r\phi^2\rangle = \frac{1}{2}\int d^3pf(\vec{p};\vec{p}_0)\sum_{l~l'}\left(F_l(r,p)\partial_rF_{l'}^*(r,p)+{\rm{cc.}}\right)P_{l}(\cos\theta)P_{l'}(\cos\theta)
\end{equation}
The product of Legendre polynomials can be simplified using the substitution, following from the identity Eq.~\eqref{eq:wigner_triple}:
\begin{equation}
    P_l(\cos\theta)P_{l'}(\cos\theta)=\sum_n(2n+1)\left\{\begin{matrix}
        l & l' & n\\0 & 0 & 0
    \end{matrix}\right\}^2P_n(\cos\theta)=\sum_nM_{ll'n}P_n(\cos\theta)
\end{equation}
yielding:
\begin{equation}
    \langle\partial_r\phi^2\rangle = \frac{1}{2}\int dp ~d{\cos\theta_p} ~d{\psi_p}~p^2 f(\vec{p};\vec{p}_0)\sum_{ll'n}M_{ll'n}\left(F_l(r,p)\partial_rF_{l'}^*(r,p)+{\rm{cc.}}\right)P_{n}(\cos\theta)
    \label{eq:wind_subbed_matrix}
\end{equation}
where the measure has additionally been expanded in spherical polar co-ordinates. Given the definition of $\cos\theta$ given in Sec.~\ref{sec:DM_wind}, the Spherical Harmonic Addition Theorem can be applied to evaluate the integral over the angle $\psi_p$:
\begin{equation}
    \int d\psi_p ~ P_n({\cos\theta}) = 2\pi P_n({\cos\theta_p})P_n({\cos\theta_r})\;,
\end{equation}
such that Eq.~\eqref{eq:wind_subbed_matrix} becomes.
\begin{equation}
    \langle\partial_r\phi^2\rangle = \pi\int dp ~d{\cos\theta_p} ~p^2 f(\vec{p};\vec{p}_0)\sum_{ll'n}M_{ll'n}\left(F_l(r,p)\partial_rF_{l'}^*(r,p)+{\rm{cc.}}\right) P_n({\cos\theta_p})P_n({\cos\theta_r})\;.
    \label{eq:wind_phi_integrated_out}
\end{equation}
The integral over $\cos\theta_p$ can then be evaluated exactly. We choose the co-ordinate system in $p$ such that $p_0$ points along the vector $\theta_p=0$. Then:
\begin{equation}
    f(\vec{p},\vec{p}_0)=\frac{1}{(\pi p_0)^{3/2}}e^{-\frac{p^2-1}{p_0^2}}e^{\frac{p}{p_0}\cos\theta_p}\;.
\end{equation}
The $\cos\theta_p$ dependent part of the integral then amounts to:
\begin{equation}
    \int_{-1}^{1} d\cos\theta_p~e^{\frac{p}{p_0}\cos\theta_p}P_n(\cos\theta_p)=2i_n\left(\frac{2p}{p_0}\right)\;,
\end{equation}
where $i_n$ are modified spherical Bessel functions of the first kind. Then we are left with an integral which is too difficult to compute analytically:
\begin{equation}
    \langle\partial_r\phi^2\rangle = \frac{2\pi}{(\pi p_0)^{3/2}}\int_0^\infty dp p^2 e^{-\frac{p^2-1}{p_0^2}}\sum_{ll'n}M_{ll'n}\left(F_l(r,p)\partial_rF_{l'}^*(r,p)+{\rm{cc.}}\right) i_n\left(\frac{2p}{p_0}\right)P_n({\cos\theta_r})\;.
\end{equation}
Averaging over the angular co-ordinate $\theta_r$, only the $n=0$ component survives. Substituting $M_{ll'0}=\frac{1}{2l+1}\delta_{ll'}$, and inserting the co-ordinate $r=1.9R_\oplus$ yields the expression given in Eq.~\eqref{eq:wind_field_gradient}. 

In regard to the computation of Eq.~\eqref{eq:wind_field_gradient}, the integrand can be simplified by neglecting contributions of the form $\sim j_l\partial_rj_l$ from the $F_l\partial_rF_l'^*+\rm{cc.}$, as these may be simplified into a term $\sim\partial_r|e^{-ipr\cos\theta}|^2$ via Eq.~\eqref{eq:exp_identity}, which is trivially zero. We choose the limit of $l$  to be $\max(1.5~pR_\oplus, 10)$ based on the observed convergent properties of Eq.~\eqref{eq:exp_identity}. We integrate up to a maximum value of $p=5p_0$ as to provide sufficient coverage of the distribution of momenta. A similar calculation is used to produce each panel in Fig.~\ref{fig:wind_profiles}.
%In integration, the size of steps should be guided by the minimum of the characteristic scales of the integrand: $p_0, \frac{1}{1.9R_\oplus}$. Within the sum over terms $F_l\partial_rF_l'^*+\rm{cc.}$ there exists a component which scales as $\sim (2l+1)(2l'+1)(-i)^{l-l'}j_l\partial_r j_l'+\rm cc.$. Such a term ultimately derives from a contribution to the derivative of the field-squared of the form: $\partial_r|e^{-ipr\cos\theta}|^2$, and can hence be ignored. Whilst the sum over $l$ should in principle to be infinite, we find it sufficient

\end{document}